\documentclass[aps,nofootinbib,showpacs,showkeys,preprinttightenlines,preprintnumbers,superscriptaddress] {article}
\usepackage{jheppub}

\usepackage{epsf,epsfig,subfigure,graphicx,amsmath,amssymb}
\usepackage{color}

\newcommand{\dis}[1]{\begin{equation}\begin{split}#1\end{split}\end{equation}}
\newcommand{\eq}[1]{Eq.~(\ref{#1})}
\newcommand{\bfrac}[2]{{\left(\frac{#1}{#2} \right)  }}

\newcommand{\VEV}[1]{\langle #1 \rangle}

\newcommand\lsim{\mathrel{\rlap{\lower4pt\hbox{\hskip1pt$\sim$}}
    \raise1pt\hbox{$<$}}}
\newcommand\gsim{\mathrel{\rlap{\lower4pt\hbox{\hskip1pt$\sim$}}
    \raise1pt\hbox{$>$}}}

\newcommand\mplanck{M_{\rm P}}
\newcommand\ie{{\it i.e.}~}
              
\newcommand\mnlsp{m_{\rm NLSP}}

\newcommand\tanb{\tan\beta}

\newcommand\squark{\widetilde q}        \newcommand\msquark{m_{\squark}}

\newcommand{\stau}{{\tilde \tau}}

\newcommand\gluino{\tilde g}
\newcommand\mgluino{m_{\gluino}}

\newcommand\second{\,{\rm sec}}

\newcommand\treh{T_{\rm R}}

\newcommand\fa{f_{a}}

\newcommand\tev{\,{\rm TeV}}
\newcommand\gev{\,{\rm GeV}}
\newcommand\mev{\,{\rm MeV}}
\newcommand\kev{\,{\rm keV}}

\newcommand\axino{{\tilde{a}}}
\newcommand\maxino{{m_{\axino}}}

\newcommand\abunda{\Omega_{\axino}h^2}

\newcommand\omegaantp{\Omega^{\rm NTP}_{\axino}}

\newcommand\sigmabar{{\overline{\sigma}}}

\newcommand\Qs{Q_{\sigma}}
\newcommand\ra{\rightarrow}
\newcommand{\mg}{m_{\rm eff}}
\newcommand{\logsoms}{\log\left(s/m_{\rm eff}^2\right)}

\newcommand\uonepq{U(1)_{\rm PQ}}

\begin{document}


\title{Axino Cold Dark Matter Revisited}

\author[a,b]{Ki-Young Choi}
\author[c]{Laura Covi}
\author[d,e]{Jihn E. Kim}
\author[f,g]{Leszek Roszkowski}

\affiliation[a]{ Asia Pacific Center for Theoretical Physics, Pohang, Gyeongbuk 790-784, Republic of Korea, and\\
Department of Physics, POSTECH, Pohang, Gyeongbuk 790-784, Republic of Korea}

\affiliation[b]{Department of Physics, Pusan National University, Busan 609-735, Korea}

\affiliation[c]{Institute for Theoretical Physics, G\"ottingen University, 37077 G\"ottingen, Germany}

\affiliation[d]{ Department of Physics and Astronomy and Center for
  Theoretical Physics, Seoul National University, Seoul 151-747,
  Korea} 
\affiliation[e]{ GIST College, Gwangju Institute of Science and Technology, Gwangju 500-712, Korea }

\affiliation[f]{National Centre for Nuclear Research, Ho{\. z}a
  69, 00-681 Warsaw, Poland}
\affiliation[g]{Department of Physics and
Astronomy, University of Sheffield, Sheffield, S3 7RH, UK}

\emailAdd{kiyoung.choi@apctp.org}
\emailAdd{Laura.Covi@theorie.physik.uni-goettingen.de}
\emailAdd{jekim@ctp.snu.ac.kr}
\emailAdd{L.Roszkowski@sheffield.ac.uk}

\color{black} \abstract{ Axino arises in supersymmetric versions of
  axion models and is a natural candidate for cold or warm dark
  matter.  Here we revisit axino dark matter produced thermally and
  non-thermally in light of recent developments.  First we discuss the
  definition of axino relative to low energy axion one for
 several KSVZ and DFSZ models of the axion.
    Then we review and refine the
    computation of the dominant QCD production in order to avoid
    unphysical cross-sections and, depending on the model, to include
    production via $SU(2) $ and $U(1)$ interactions and Yukawa couplings.
}


\arxivnumber{1108.2282}

\maketitle

\section{Introduction}\label{sect:intro}
The lightest supersymmetric particle (LSP) with R-parity conservation
is absolutely stable and can contribute to the present Universe energy
density as a dominant component of cold dark matter (CDM).  If the
strong CP problem is solved by introducing a light pseudoscalar, an axion, its fermionic SUSY partner,
an {\em axino}, can be a natural candidate for CDM if it is the LSP.
The relic axino CDM can be produced either non-thermally, through
out-of-equilibrium decays, or thermally, through scatterings and
decays in the hot plasma, as originally shown in
Refs.~\cite{CKR00,CKKR01}. The scenario was subsequently extensively studied during
the last decade~\cite{CRS02,CRRS04,Steffen04,Strumia10,AxinoRevs,flaxino,
  Baer, Wyler09}, with either a neutralino or a stau as the
next-to-lightest supersymmetric particle (NLSP). Ways of testing the
axino CDM scenario at the LHC were explored
in Refs.~\cite{Brandenburg05,Hamaguchi:2006vu,ChoiKY07,Wyler09}  and implications for Affleck-Dine
Baryogenesis in Ref.~\cite{Roszkowski:2006kw}.

The strong CP problem is naturally solved by introducing a very light
axion field $a$. The axion appears when the Peccei-Quinn (PQ) symmetry is
broken at some scale $\fa$. Below the PQ scale, after integrating out
heavy quarks carrying PQ charges~\cite{KSVZ79}, an effective
axion--gluon interaction is given by
\dis{
{\mathcal L}_a^{\rm eff}=\frac{\alpha_s}{8\pi \fa}a\, G_{\mu\nu} \widetilde{G}^{\mu\nu},
\label{Leffa}
}
where $\alpha_s$ is the strong coupling constant, $G$ is the field
strength of the gluon field and $\widetilde{G}_{\mu\nu}\equiv
\frac12 \epsilon_{\mu\nu\rho\sigma} G^{\mu\nu}$ is its dual with $\epsilon_{0123}=-1$.
Different types of axion models have been proposed, with distinctively
different couplings to Standard Model (SM) fields, depending
on their PQ charge assignment.

Very light axion models contain a complex SM singlet scalar field
carrying a PQ charge.  In the Kim-Shifman-Vainstein-Zakharov~(KSVZ)
class of models~\cite{KSVZ79} the PQ charges are assigned to new heavy
quarks, while in the Dine-Fischler-Srednicki-Zhitnitskii~(DFSZ)
approach~\cite{DFSZ81} the PQ charges are assigned to the SM
quarks. This difference is the origin of different phenomenological
properties~\cite{KimRMP10} of the KSVZ and DFSZ classes of models
since in the low energy effective theory at the electroweak (EW) scale
(after integrating out heavy fields) the gluon anomaly term is the
source of all interactions in the KSVZ models while the Yukawa
couplings are the source of all interactions in the DFSZ models.  For
solving the strong CP problem, one needs a coupling of the axion to
the gluon anomaly and this is generated by a heavy quark loop in the
KVSZ models, appearing as a non-renormalizable effective coupling when
those heavy fields are integrated out. In the DFSZ models instead the
coupling is generated by SM quark loops.

The couplings arising in these two popular classes of models will be discussed in the
next section. For the KSVZ axion, constraints on the PQ scale $\fa$
have been obtained from astrophysical and cosmological considerations
and the scale is limited to a rather narrow window $10^{10}\gev
\lsim \fa \lsim 10^{12}\gev$~\cite{KimRMP10}, while for the
DFSZ case precise constraints on $\fa$ have not been derived yet.

The axino as a candidate for CDM was originally studied mostly for the SUSY
version of the KSVZ axion model, in an important production mode
corresponding to the interaction term given by \eq{Leffa}~\cite{CKKR01}.
The supersymmetrization of axion models
introduces a full axion supermultiplet $A$ which contains the
pseudoscalar axion $a$, its scalar partner {\em saxion} $s$, and their
fermionic partner axino $\axino$,
\dis{
  A=\frac{1}{\sqrt2}(s+ia)+\sqrt2 \axino \vartheta + F_A
  \vartheta\vartheta,
}
where $F_A$ stands for an auxiliary field and $\vartheta$ for a Grassmann coordinate.

The effective axion interaction of \eq{Leffa}
can  easily be supersymmetrized using the superpotential of the axion
and the vector multiplet $W_\alpha$ containing the gluon field
\dis{ {\mathcal
    L}^{\rm eff} =-\frac{\alpha_s}{2\sqrt2\pi \fa}\int A\,{\rm
    Tr}\,[W_\alpha W^\alpha].
\label{Leff3}
}
Effective axion multiplet interactions with the other gauge bosons can
be obtained in a similar way.
Axino production from QCD scatterings due to interaction~\eq{Leff3} has been
considered in the literature in different approximations, which have
led to somewhat different numerical results. The main technical
difficulty and source of uncertainty is the question of how to
regulate the infrared divergences due to the exchange of massless gluons.
In the original study~\cite{CKKR01} a simple insertion of a thermal
gluon mass to regulate infrared (IR) divergences was used and the
leading logarithmic term was obtained, without much control over the subleading
finite piece, since the thermal mass was introduced by hand.
Subsequently, a hard thermal loop (HTL) resummation method was applied
in Ref.~\cite{Steffen04}, allowing for a self-consistent determination of
the gluon thermal mass and more control over the constant term in the
high energy region of axino production.
Recently, in Ref.~\cite{Strumia10} a new calculation was presented
which, although not  gauge invariant, includes more terms
of the perturbative series, in particular the decay of the gluon
whose thermal mass can be larger than the gluino and axino mass taken together.
The two latter methods have their own advantages and limitations, and
they coincide in the high energy region where the convergence
of the perturbation series is stable.
In Ref.~\cite{Strumia10} a previously neglected dimension-5
term in the Lagrangian, containing purely interactions between supersymmetric
particles was also included. This however changes the axino production rate by less
than 1\%. In our paper, we adopt the original way of effective mass approximation
but we improve it to make the result positive definite. Although this
method is not gauge invariant either,
it is only known viable method at relatively low reheating
temperatures, which is the regime important for the axino
as cold dark matter candidate.
We shall come back to this discussion in more detail below.

In the calculations published so far the couplings of the axino to the gauge
multiplets other than the one of the gluon was also neglected.
In fact, in Ref.~\cite{CKKR01} a chiral transformation of the
left-handed lepton doublets was performed to remove the axion $SU(2)$
anomaly interaction and to leave only the axion $U(1)$ contribution.
Then the axion $SU(2)$ anomaly coupling re-appears in
principle from the leptonic loops, which are independent
of the fermion masses.
The corresponding axino loops, on the other hand, are suppressed by
the ratio $m_{\rm lepton}^2/M^2$ where $M$ is the largest mass
in the loop. Therefore the error in neglecting the axion $SU(2)$
anomaly is estimated to be at most of order $m_\tau^2/m_{\tilde\tau}^2$
compared to the $SU(3)$ term.
However, in supersymmetric extensions of axionic models, the chiral
rotation on the lepton fields involves also their scalar counterparts,
the sleptons and the saxion, which generates additional
couplings for those fields. It is also not clear if it is justified
to perform a redefinition in a fully supersymmetric way when
supersymmetry is in any case broken. Rather than going into the swamp
of such interactions with unknown slepton and saxion masses as
parameters, we will keep here the general axion $SU(2)$ anomaly
interaction and its SUSY counterpart, which is more tractable, and
examine its effect on the axino abundance.

Moreover, in the present work we will also consider the role of
Yukawa-type interactions between the axino and the matter multiplets,
which arise either at one loop or at tree level, depending on the
model.  We will investigate what effect these terms may have and in
particular how model dependent our results for the axino abundance
are.  In the previous study, only the KSVZ model was considered for
the axino production.  However in the DFSZ model axino production is
dominated by the Yukawa coupling and the dependence on the reheating
temperature is quite different from that in the KSVZ model.
Only recently Refs.~\cite{Chun:2011zd} and~\cite{Bae:2011jb}
  considered axinos in the DFSZ model.  In particular,
  Ref.~\cite{Chun:2011zd} gave a simple approximate formulae for the
  relic density of light axino as dark matter.  In our paper, we study
  axino production in the DFSZ model and the suitability of the DFSZ
  axino as dark matter in a more complete way. 

In view of the recent developments in estimating the QCD contribution,
in this paper we also update and re-examine
relic CDM axino production, with an emphasis on an estimate of
the uncertainties as well as on model dependence.
In  particular, our updated analysis of the axino CDM scenario
improves on the previous works in the
following aspects:\\
(i) an inclusion of some previously neglected terms in the axino production
 and of subleading terms in the mass of the gluon
beyond the logarithmic and constant pieces -- these last parts
ensure that the cross-section remains positive even for the invariant
mass $s$ smaller than the gluon thermal mass; \\
(ii) an explicit calculation of axino production via $SU(2)$ and $U(1)$
interactions; \\
(iii) a derivation of the axino abundance in specific implementations
of the KSVZ and DFSZ models. \\
(iv) an update on the constraints on the reheating temperature
$\treh$~\footnote{ We assume the instant reheating approximation
    and define the reheating temperature as the maximum temperature at
    which standard Big Bang expansion with a thermalised bath of SM
    particles starts. We can easily translate the axino abundance
    given with this conventional definition of the reheating
    temperature to more specific reheating scenarios. E.g.,
    Ref.~\cite{Strumia10} considers a reheating process from the decay
    of the inflaton and obtains a slightly smaller abundance of
    axinos, reduced by a factor of 0.745.  In comparing our results
    we account for this difference. }  for the both the
neutralino and the stau as the NLSP using the current WMAP-7 result on
the DM relic density and relevant structure formation data.

In Sect.~\ref{sec:AxinoTh}, we define the axino by its relation to the
shift symmetry of the KSVZ and DFSZ axions. In
Sect.~\ref{sec:AxinoProd} we calculate axino production rate for the
KSVZ and DFSZ axion models, and in Sect.~\ref{sec:CosBounds} we
discuss several constraints on the scenario arising from cosmology.
Finally, in Sect.~\ref{sec:Concl} we present our
conclusions.

\section{Axinos}\label{sec:AxinoTh}

\subsection{The framework}

In a recent review~\cite{KimRMP10} low energy axion interactions were given in terms
of the effective couplings with the SM fields $c_1$, $c_2,$ and
$c_3$, which arise after integrating out all heavy PQ charged fields.
The effective axion Lagrangian terms are
\dis{
&\frac{(\partial_\mu a)}{\fa}\sum_i\left(c_1^u\bar u_i
  \gamma^\mu\gamma_5 u_i+c_1^d \bar d_i \gamma^\mu\gamma_5
  d_i\right)\\
&+\frac{a}{\fa}\sum_i\left(c_2^um_u^i\bar u_i i\gamma_5 u_i+
 c_2^dm_d^i\bar d_i i\gamma_5 d_i\right)\\
 &\quad\quad\quad+\frac{c_3}{32\pi^2 \fa}
 a G\tilde G, \label{eq:efflagr}
}
where $c_3$ can be defined to be 1 (if it is non-zero) by rescaling
$\fa$.\footnote{The strong coupling $g_3$ is usually omitted by
  absorbing it to the field strengths. Here, $\fa$ denotes the so-called
  axion decay constant which is typically related to the vacuum
  expectation value $V$ of the PQ symmetry breaking scalar field by
  $\fa=V/N$ where $N$ is the domain wall number.}
The $c_1 $ term is the PQ symmetry preserving derivative
interaction of the axion field that can be reabsorbed into the $c_2 $
term by a partial integration over on-shell quarks. For the $c_2$
terms, we have only kept the lowest order terms (proportional to
$1/\fa$), while in principle an infinite series of terms in $ a/\fa $
arises.

In the following we will consider the two popular scenarios mentioned earlier: the
KSVZ and the DFSZ classes of axion models.  The KSVZ class of axion models corresponds to
the choice $c_1=c_2=0, c_3=1$, and the DFSZ one to $c_1=c_3=0$
and $c_2\ne 0$, after integrating out the heavy field sector responsible
for PQ symmetry breaking.  In the
latter model, if the Higgs doublets $H_{u,d}$ carry respective PQ
charges $ Q_{u,d}$, the SM fields also carry PQ charges,\footnote{In
  variant axion models, $ Q_{u,d}$ may have family dependence but
  here we suppress family indices. They can be inserted if
  needed.}  see Table~I, and the anomaly interaction proportional to
$c_3 \neq 0 $ arises from SM quark loops.  The axion mass is given by
the strong interaction and is proportional to $|c_2+c_3|$. Hence the
sum $c_2+c_3$, if it is non-zero, defines the QCD axion. Only this
combination of the two couplings is physical, since a chiral
axion-dependent PQ rotation of the quark fields can shift the values
of $c_2 $ and $c_3$, while keeping $c_2+c_3$ constant.  This is connected to
the well-known fact that, if one of the quark masses is zero then the
anomaly becomes unphysical and can be reabsorbed in the rotation of
the massless field.

In the supersymmetric version of axionic models, the interactions of
the saxion and the axino with matter are related by supersymmetry to
those of the axion. Hence the definition of the axion at low energy must be
connected to the definition of the axion multiplet, and therefore of
the saxion and the axino, at energies above the EW scale.\footnote{The
  supersymmetrization of axion models was first discussed in
  Refs.~\cite{Tamv82,NillesRaby82,Frere83}. An explicit model was
  first constructed in~\cite{Kim83}, and the first cosmological study
  was performed in Ref.~\cite{Masiero84}.}

Below we will examine more closely the KSVZ and DFSZ models of the
axion. In both models one imposes the PQ symmetry at a high energy
scale and the axion emerges from its spontaneous symmetry breaking.
As a  specific example, let us consider the PQ sector at high energy with the
PQ breaking implemented as in Ref.~\cite{Kim83} \dis{ W_{\rm PQ}=f_Z
  Z(S_1S_2-V_a^2),\label{KimW} } where $Z,S_1$, and $S_2$ are gauge
singlet chiral superfields, $f_Z$ is a Yukawa coupling and $V_a$ is a
parameter in the Lagrangian which determines non-zero VEVs
$\langle S_1 \rangle= \langle S_2\rangle =V_a$ in
the minimization process.  The superfields transform under the
$\uonepq$ symmetry as \dis{ Z\rightarrow Z,\quad S_1 \rightarrow
  e^{i\alpha\Qs}S_1,\quad S_2 \rightarrow e^{-i\alpha\Qs}S_2.  }
\begin{table}
\begin{center}
\begin{tabular}{|c|c|c|c|c|c|c|c|c|c|c|}
\hline
Model  & $Z$ & $S_1$ & $S_2$ & $Q_L$ & $\overline{Q}_R$
&$H_d$ &$H_u$& $q_L$& $D^c_R$ & $U^c_R$ \\
\hline
{\rm KSVZ}& 0 & $\Qs$&$-\Qs$&$-\frac12\Qs$&$-\frac12\Qs$& 0& 0& 0&0&0\\
{\rm DFSZ}& 0 & $\Qs$&$-\Qs$& 0& 0&$-Q_d$ & $-Q_u$& 0& $+Q_d$ &$+Q_u$\\
\hline
\end{tabular}
\caption{The PQ charge assignment $Q$. $Q_L$ and $\overline{Q}_R$ denote
new heavy  quark multiplets. In the DFSZ model $2\Qs=Q_u+Q_d$, and the
  PQ charge of the left-handed SM quark doublets $q_L$
  vanishes. See text for more details. }
\label{table:PQcharge}
\end{center}
\end{table}
The potential in the global supersymmetric limit is
\dis{ V=\sum_a
  \left|\frac{\partial W}{\partial \phi_a} \right|^2 + \frac12 D^a
  D^a,
}
with
\dis{ D^a=g\sum_\phi \phi^*T^a\phi,
}
where $g$ is the
gauge coupling of the gauge groups and $\phi$ denotes collectively all
the scalar fields.  With the superpotential given by
\eq{KimW} one has
\dis{
  &\frac{\partial W}{\partial Z} = f_Z(S_1S_2-V_a^2),\\
  &\frac{\partial W}{\partial S_1} =f_Z Z S_2,\\
  &\frac{\partial W}{\partial S_2} =f_Z Z S_1.\label{eq:Kim83} }
At the
tree-level both $S_1$ and $S_2$ develop VEVs and break the PQ
symmetry.  The fermionic partners of $Z$ and $S'=(S_1+S_2)/\sqrt2$
combine to become a Dirac fermion with mass $m_Z=\sqrt2 f_Z V_a $, while
$S=(S_1-S_2)/\sqrt2$  contains the axion field and can be
identified with the axion multiplet.  From \eq{eq:Kim83} we note that,
if $\langle Z\rangle =0$ and SUSY is not broken then both the axino
and the saxion are mass degenerate with the axion.

However, with soft
SUSY breaking terms included, $\langle Z\rangle$ can develop a non-zero
value, in which case both the saxion and the axino become massive, independently of
the axion.
Therefore, a full specification of the SUSY breaking mechanism is
needed in order to determine the mass of the axino exactly. This was first studied
in Ref.~\cite{Kim83} for the superpotential above, while another
superpotential and SUSY breaking with a very small axino mass was
given in Ref.~\cite{ChunKN92}.
Recently the case of a direct coupling between the axion and
SUSY breaking sector was discussed in  Ref.~\cite{Higaki:2011bz}.

\subsection{The KSVZ model}
In the KSVZ approach in order to obtain the anomalous interaction of the axion and the
gluon fields, one introduces the heavy quark fields $Q_L$ and $Q_R$ in the superpotential as Ref.~\cite{KSVZ79}
\dis{
W_{\rm KSVZ}=W_{\rm PQ}+f_Q Q_L\overline{Q}_R S_1.\label{Wksvz}
}
After the $\uonepq$ symmetry breaking takes place, as discussed above, $Q_L$ and $Q_R$
combine to become a heavy Dirac fermion with mass $m_Q=f_Q V_a$,
Assuming that $ V_a \gg m_{3/2} $, the scalar partner has practically the
same mass and the whole supermultiplet can be integrated out in a
supersymmetric way.
The low-energy Lagrangian can then be obtained by integrating out the heavy
quark multiplet or by using anomaly matching condition.
In this way one finds that the axion anomaly term becomes
\dis{
{\mathcal L}^{\rm eff}
= \frac{ \alpha_s
  N_Q}{8\pi \fa }a\, G^{\mu\nu}\widetilde{G}_{\mu\nu},\label{axion}
}
where the axion $a$ is the pseudoscalar component of the superfield $S\equiv
(S_1-S_2)/\sqrt2$, and the axion decay constant is  $\fa=2V_a$. $N_Q$
is the number of the heavy quarks and we consider $N_Q=1$ in our
case. Then we have $c_1=c_2=0$ and $c_3=1$~\cite{KimRMP10}.

The low-energy interactions of the saxion and the axino fields can be obtained in the same
way by integrating out the heavy (s)quark fields. However, in the
limit of unbroken SUSY the low-energy effective Lagrangian should be
given in a SUSY invariant form, including \eq{axion}, as
\dis{
{\mathcal L}^{\rm eff}=-\frac{\alpha_s}{2\sqrt{2} \pi \fa}\int
d^2\,\vartheta A\, W^{a  \alpha} W^a_{\alpha} +
\textrm{h.c.} \label{LeffKSVZ}
}

The axion superfield $A$ and  $W^{a\,\alpha}$ are given by~\cite{WessBag92}
\dis{
&A=\frac{s+ia}{\sqrt2}+\sqrt2 \vartheta{\psi_a} + \vartheta^2 F_A,\\
&W^{a}_{\alpha}=-i{\lambda}^{a}_{\alpha} + [\delta_\alpha^\beta D^a
 -\frac{i}{2}(\sigma^{\mu}\bar\sigma^\nu)_\alpha^\beta G_{\mu\nu}^a]\vartheta_\beta
+\vartheta\vartheta\sigma^\mu_{\alpha\dot{\alpha}} D_\mu
\bar{{\lambda}}^{\dot{\alpha}}.
\label{component}
}
The effective Lagrangian in terms of the Bjorken-Drell gamma matrices
then reads~\footnote{Our saxion-gluino-gluino interaction
    differs by a factor of 2 from that in Ref.~\cite{Strumia10}.}.
\dis{
{\mathcal L}^{\rm eff}=& \frac{\alpha_s}{8\pi \fa}\left[
  a(G^{a\mu\nu}\widetilde{G}^a_{\mu\nu}
  +D_\mu(\overline{\tilde{g}}\gamma^\mu\gamma^5\tilde{g}) ) \right.
  \\
&+s(G^{a\mu\nu}G^a_{\mu\nu}-2D^aD^a+2i \overline{\tilde{g}}\gamma^\mu D_\mu\tilde{g})\\
&\left. + i\overline{\axino}G^{a}_{\mu\nu}\frac{[\gamma^\mu,\gamma^\nu]}{2}
\gamma^5\gluino-2 \overline{\axino}\gluino D^a\right]\\
& +\sqrt2\left(F_A\lambda\lambda + \textrm{h.c.}  \right). ,
\label{eq:Leff}
}
where the gluino and axino 4-spinors are given by
\dis{
\tilde{g}=\left( \begin{array}{c}
-i\lambda\\
i\bar{\lambda}
\end{array}\right),\qquad
\tilde{a}=\left( \begin{array}{c}
\psi_a\\
\overline{\psi}_a
\end{array}\right).
}
The effective Lagrangian~(\ref{eq:Leff}), including the axino interaction
with the D-term has recently been used in Ref.~\cite{Strumia10} to calculate the thermal
production rate of axino dark matter and will be the basis also for the
present analysis.

\subsection{The DFSZ model}
In the DFSZ framework, the $SU(2)_L\times U(1)_Y$ Higgs doublets carry PQ charges
and thus the light quarks are also charged under $\uonepq$.
The charge assignment is shown in Table~\ref{table:PQcharge}.
The anomaly coupling $a G\widetilde{G}$ can be obtained after electroweak
symmetry breaking (EWSB) through the coupling
of the axion to the Higgs doublets which couple to the light quarks.
To this end, one adds a non-renormalizable term to the PQ breaking
superpotential~\cite{Kim:1983dt}
\dis{
 W_{\rm DFSZ}=W_{\rm PQ}+ \frac{f_s}{\mplanck} S_1^2 H_d H_u,
\label{eq:DFZSsuperpotential}
}
where $W_{\rm PQ}$ is given in \eq{KimW} and
$H_d H_u\equiv \epsilon_{\alpha\beta}H_d^\alpha H_u^\beta$.
Note that here $ f_s \sim \mu M_P/V_a^2 \sim 1 $
generates a phenomenologically acceptable supersymmetric $ \mu$-term.

The superpotential including light quarks is given by
\dis{
W_{\rm MSSM}=y_t Q H_u U^c +y_b Q H_d D^c +y_\tau L H_d E^c.
\label{WMSSM}
}

Before the EW symmetry is broken but after the
$\uonepq$ symmetry is broken, the massless axion Goldstone boson is identified as
\dis{
a=\frac{a_1-a_2}{\sqrt2}.\label{axion_DFSZ_high}
}
Here we defined the component fields in the same way as in
   \eq{component},
   \dis{
   S_1=\frac{s_1+ia_1}{\sqrt2}+\sqrt2 \vartheta \tilde{a}_1 + \vartheta^2 F_1,
   }
   and similarly for $S_2$.  
Instead, the axino mass eigenstate can be obtained from the mass matrix
\dis{
f_Z\begin{bmatrix}
0& z_0 & \VEV{S_2} \\
z_0&0&\VEV{S_1}\\
\VEV{S_2}& \VEV{S_1}& 0
\end{bmatrix}
\simeq
f_Z\begin{bmatrix}
0& z_0 & V_a \\
z_0&0&V_a\\
V_a& V_a& 0
\end{bmatrix}.  } in the basis of
$(\tilde{a}_1,\tilde{a}_2,\tilde{Z})$ and we have used $z_0=\VEV{Z}$.
The lightest state, which we will identify with the axino, has mass
$f_Zz_0$ and is given by \dis{ \tilde{a}=&
  \frac{\tilde{a}_1-\tilde{a}_2}{\sqrt2}, } which coincides
  with \eq{axion_DFSZ_high}.      Since EW symmetry is not broken, the
axion (and thus the axino) do not mix with the Higgs (and higgsino)
and therefore the axion cannot have the $SU(3)_c$ anomalous
interaction with gluons generated at one loop via the quark triangle
diagrams.\footnote{An axino-gluino-gluon coupling can arise at two
  loops through the Higgs Yukawa couplings, which are non-vanishing
  even above EW symmetry breaking, but it is strongly suppressed and
  we will neglect it here.}  However, $SU(2)_L$ and $U(1)_Y$ anomalous
interaction can be generated via a higgsino triangle loop through the
Yukawa coupling derived from \eq{eq:DFZSsuperpotential}.  They are
given by
\dis{ {\mathcal L}_{\rm anomaly}= -\frac{\alpha_2}{8\pi
    V_a}aW^a_{\mu\nu}\widetilde{W}^{a\mu\nu}-\frac{\alpha_1}{8\pi
    V_a}aB_{\mu\nu}\widetilde{B}^{\mu\nu}.
\label{DFSZanomaly21}
}
These anomaly interactions appear also for the axino, but
may obtain corrections of order of one from SUSY breaking,
compared to the axion couplings, and such uncertainties will be
later encoded in the coefficients $C_{aWW}, C_{aYY} $.

After the EW and the $\uonepq$ symmetries are both broken,
the axion is identified with the Goldstone boson of the broken $\uonepq$
symmetry, given by
\dis{
a = \frac{2\Qs V_a a_s-Q_dv_d P_d -Q_uv_uP_u}{\sqrt{4 \Qs^2V_a^2 +Q_d^2v_d^2+Q_u^2v_u^2}},\label{eq:axiondef}
}
where $a_s=(a_1-a_2)/\sqrt2$, and we expanded the Higgs field as
\dis{
H_d^0= &\frac{v_d+h_d}{\sqrt2}e^{iP_d/v_d},\qquad
H_u^0= \frac{v_u+h_u}{\sqrt2}e^{iP_u/v_u},\\
&\qquad v=\sqrt{v_d^2+v_u^2}\,.
}
with $v_d/\sqrt{2}=\langle 0|H_d|0\rangle$ and
$v_u/\sqrt{2}=\langle 0|H_u|0\rangle$, while $ P_{d,u} $ are the pseudoscalar
fields contained in the electrically neutral component of $ H_{d,u}
$.

The neutral Higgs boson component eaten by the $Z$-boson and the orthogonal
pseudoscalar Higgs, \ie the phases of $H_d,H_u$ and $S_1-S_2 $, are
given by
\dis{
&Z^L=\frac{-v_dP_d+v_uP_u}{v},\\
&A=\frac{v_dv_ua_s+V_a v_uP_d +\fa v_dP_u}{\sqrt{V_a^2v^2 +v_d^2v_u^2}}\\
&a = \frac{(\frac{v_d}{v_u}+\frac{v_u}{v_d})V_a a_s-v_uP_d -v_dP_u}{v \sqrt{v^2/(v_d^2v_u^2)V_a^2 + 1}}\\
&\qquad = \frac{v^2V_a a_s-v_dv_u^2P_d -v_uv_d^2P_u}{v\sqrt{v^2V_a^2 +v_d^2v_u^2}}
\,.\label{eq:pseudorth}
}
Equating (\ref{eq:axiondef}) with the last term of \eq{eq:pseudorth},
we obtain
\dis{
&Q_d =\frac{v_u}{v_d} = \tan\beta ,\qquad
Q_u=\frac{v_d}{v_u} = 1/\tan\beta,\\
& Q_\sigma= \frac12\left(\frac{v_d}{v_u}+\frac{v_u}{v_d} \right)
= \frac{1}{2 \sin\beta\cos\beta},
}
up to a common normalization constant.

The interactions of the axion with the matter fields are obtained through the
axion part of the phase of the Higgs fields,
\dis{
&P_d \simeq \frac{v_dv_u^2}{v^2V_a}a + \cdots = \frac{v}{V_a} \sin^2\beta \cos\beta + \cdots,\\
&P_u \simeq \frac{v_d^2v_u}{v^2V_a}a + \cdots = \frac{v}{V_a} \sin\beta \cos^2\beta + \cdots.
}
Considering the Yukawa interaction from the superpotential, \eq{WMSSM},
\dis{
{\mathcal L}=-y_t \overline{u}_Ru_LH_u^0-y_t^*\overline{u}_Lu_RH_u^{0\,*}+\cdots,
}
the Lagrangian terms for the up-type quark axion couplings are given by
\dis{
{\mathcal L}= &-y_t \frac{v_u}{\sqrt2}\overline{u}_Ru_L\exp\left[i\frac{v_d^2}{v^2}
\frac{a}{V_a} \right]\\
&-y_t^*\frac{v_u}{\sqrt2}\overline{u}_Lu_R\exp
\left[-i\frac{v_d^2}{v^2}\frac{a}{V_a} \right]+\cdots, \label{eq:DFSZYuk}
}
and similarly for the down-type quarks.
We can now compare the above to the coefficients in the definition of
the axion~\cite{KimRMP10} and obtain for following values for
$c_2^{u,d}$; compare \eq{eq:efflagr},
\dis{
c_2^{u}= \frac{v_d^2}{v^2} = \cos^2\beta ,\qquad
c_2^{d}= \frac{v_u^2}{v^2} = \sin^2\beta.
\label{eq:DFSZc2all}
}

After integrating out all the Higgs fields except the axion supermultiplet, which remains light,
all the axion couplings  arise from the $c_2$ terms given in \eq{eq:DFSZYuk}  and
\eq{eq:DFSZc2all} at low energies above the quark masses.
At one loop in the SM fermions one obtains the axion-anomaly
interaction term. It is then given by
\dis{
{\mathcal L}_{\rm anomaly}=\frac{\alpha_s N}{8\pi (2V_a)}aG^{\mu\nu}\widetilde{G}_{\mu\nu}\,,\label{DFSZanomaly3}
}
where $N=6$ and again $\fa=2V_a/N$.

In the supersymmetric limit, below EW symmetry breaking, the axino
mass eigenstate can be read off from \eq{eq:axiondef} and is given by
\dis{
\axino= \frac{2\Qs V_a \tilde{a}_s-Q_dv_d\tilde{h}_d
  -Q_uv_u\tilde{h}_u}{\sqrt{4\Qs^2 V_a^2
    +Q_d^2v_d^2+Q_u^2v_u^2}}\,. \label{eq:axinoDFSZ}
}
Here  $ \tilde h_{d,u} $ denote the fermionic components of the
electrically neutral parts of $ H_{d,u} $.
However since supersymmetry is broken, in general in the DFSZ models
the axino mixes with the higgsinos differently than the axion with
the Higgs and the mass eigenstate is not exactly the state given in
\eq{eq:axinoDFSZ}. Such a field is a good approximation to the
physical axino only if the mixing generated from the
superpotential~(\ref{eq:DFZSsuperpotential}) and SUSY breaking,
which is of order $v_{u,d}/\fa $ and $z_0/\fa$, is negligible. Otherwise
the whole axino-neutralino mixing matrix has to be considered in
detail.  We will not discuss this case further, but point instead to
the related studies in the case of the
Next-to-Minimal Supersymmetric Standard Model with a
singlino LSP~\cite{NMSSM}.

In the DFSZ models, there are also axino tree-level Yukawa interaction terms
to the quark and squark with a coupling of the order of $m_q/\fa$
below the EWSB scale, with the Higgs and higgsino with
coupling $\mu/\fa$ even above the EWSB scale.
These tree-level interactions are not suppressed by a gauge
coupling or a loop factor, as the QCD anomaly term has, and thus they
give the dominant contribution to axino production through the
decay and/or scattering processes at low reheating temperature $\treh$.
At high reheating temperatures above EWSB instead the $ SU(3)_c$
anomaly coupling is absent and the $SU(2)_L$ anomalous interaction
dominates the production.

\section{The production of axinos}\label{sec:AxinoProd}

There are two efficient and robust ways of populating the early
Universe with axinos. Firstly, they can be produced through
scatterings and decays of particles in thermal equilibrium. This
mechanism, which we call {\em thermal production} (TP) depends on the
reheating temperature after inflation.  The other mechanism, which is
independent of $\treh$, involves {\em non-thermal production} (NTP) of
axinos from the decay of the NLSP after it has frozen out from the
plasma.  Note also that, even though squarks are normally not the NLSP
and remain in thermal equilibrium, for $\treh \lsim \msquark$ and
large gluino mass, axino yield from  decay processes
$\squark\rightarrow q \axino$ can dominate the abundance~\cite{CRS02}.
Additionally the decay of inflaton or moduli can produce axinos but
such contributions are very model dependent and won't be considered here.

Thermally produced axinos in the $\kev$ mass range were considered as
warm dark matter (WDM) in Ref.~\cite{RTW91} and much lighter ones as hot DM
in Ref.~\cite{Masiero84}.  In Ref.~\cite{CKR00} it was shown that axinos from
neutralino NLSP decays can be a natural candidate for CDM and this was
extended in Ref.~\cite{CKKR01} to include TP.  If the axino mass is between
around an MeV to several GeV, the correct axino CDM density is obtained
when $\treh$ is less than about $5\times 10^4\gev$~\cite{CKKR01}.  At
higher $\treh$ and lower ($\sim\kev$) mass, axinos could constitute
WDM.  As a digression, we note that, when axinos are very heavy, and
it is the neutralinos that play the role of the LSP, their population
from heavy axino decays could constitute CDM~\cite{CKLS08}, leading in
particular to the possibility of TeV-scale cosmic ray positrons, as
pointed out in Ref.~\cite{HuhDecay09}. The possibility of either WDM
or very heavy axinos is not discussed in this paper.

The interactions leading to CDM axinos were extensively studied in
terms of cosmological implications in Refs.~\cite{CKKR01,CRS02,CRRS04}
and collider signatures in Refs.~\cite{Brandenburg05,Hamaguchi:2006vu,ChoiKY07,Wyler09}.
If the LHC does not confirm the decay of heavy squarks or gluino to a lighter neutralino, the axino CDM idea with the R-parity conservation can not be saved unless some other mechanisms such as an effective SUSY is introduced \cite{effSUSY}.

In general the couplings of the axino to gauge and matter fields are
analogous to those give in \eq{LeffKSVZ}
\dis{
{\mathcal L}^{\rm eff}=&-\frac{\sqrt2 \alpha_s}{8\pi \fa}\int
d^2\,\vartheta S\, G^{a  \alpha}G^a_{\alpha}\\&-\frac{\sqrt2
  \alpha_2 C_{aWW}}{8\pi \fa}\int d^2\,\vartheta S\, W^{a  \alpha}
W^a_{\alpha}\\&-\frac{\sqrt2 \alpha_Y C_{aYY}}{8\pi \fa}\int
d^2\,\vartheta S\, Y^{a  \alpha} Y^a_{\alpha} +
\textrm{h.c.}. \label{Leff}
}
Here we normalize the PQ scale by $2V_a/N\rightarrow \fa$ which
sets $ c_3 =1 $ for the QCD anomaly coupling and therefore defines the
coefficients $C_{aWW}$ and $C_{aYY}$ in the axion models considered above.

\subsection{Thermal production}

At sufficiently high temperatures ($\gsim10^9\gev$) axinos
can reach thermal equilibrium with SM particles and their
superpartners.  However, assuming that a subsequent period of cosmic
inflation dilutes the population of such primordial axinos (and that
they are not produced directly in inflaton decay), a post-inflationary
axino population comes firstly from the hot thermal bath.  If the
reheating temperature is very high, above the decoupling temperature
of axinos ($\sim10^9\gev$), their relic number density reaches again
thermal equilibrium and is the same as that of photons. In that case
axinos must be so light ( $\lsim 1\kev$) that they become warm or
hot DM~\cite{RTW91}. On the other hand, when the reheating temperature
is below the decoupling temperature, axino number density is much
smaller than that of photons and its time evolution is well described
by the Boltzmann equation without backreaction~\cite{CKKR01}.

In the KSVZ model, at high temperatures, the most important contributions come from two-body
scatterings of colored particles into an axino and other particles. At
lower reheating temperatures, on the other hand,
the decay of squarks or gluinos can dominate the production of
axinos~\cite{CRS02,CRRS04}.
In Ref.~\cite{CKKR01}, an effective gluon thermal mass was introduced
to regulate the infrared divergence in the scattering cross-section.
Subsequently, a more consistent calculation using the hard thermal loop (HTL)
approximation was applied to the axino production in Ref.~\cite{Steffen04}.
However, the HTL approximation is valid only for small gauge coupling,
$g\ll 1$, which corresponds to the reheating temperature
$\treh \gg 10^6\gev$. 
Below $10^6\gev$ the HTL approximation becomes less 
reliable~\cite{Steffen04}. In fact, the production rate 
becomes even negative at $g_3\gtrsim 1.2$, and therefore the result becomes
unphysical. Strumia tried to improve this behaviour by using the full 
resummed finite-temperature propagators for gluons and gluinos in the
loop~\cite{Strumia10}. This procedure includes axino production via 
gluon decays, which is kinematically allowed by thermal mass, and 
results in an enhancement compared to the HTL approximation.
However, his method is gauge-dependent in the next-to-leading
order~\cite{RS06}, indicating that not all the contributions of that
order are included, and therefore also does not give a fully
satisfactory result in the large gauge coupling regime. 
For these reasons, we believe that it is worth pursuing an
alternative method of computing the rate at large couplings and
we apply the effective mass approximation as the only way to evaluate
the axino thermal production at large coupling $g$ and low 
reheating temperature $\treh\lesssim 10^4\gev$. 
Even though  this method is also not gauge invariant, it captures
relevant physical effects like plasma screening and gives
positive and physical cross-sections for each single scattering
process.

As stated earlier, in the previous studies of TP of relic axinos the
contributions from $SU(2)_L$ and $U(1)_Y$ gauge interactions were
neglected,  but for completeness,
we will discuss here all
SM gauge groups explicitly in order to examine their possible effects.

To start with, in evaluating the contributions from the strong
interactions we will follow the method used previously
in Ref.~\cite{CKKR01} to obtain the  (dominant) axino TP cross-section.
We will further update and correct the tables presented there by
following Ref.~\cite{Strumia10} to include the previously ignored
dimension-5 axino-gluino-squark-squark interaction term.  This term
changes only the contribution from the processes H and J in
Table~\ref{table:channels}, while it does not affect the other terms.
The processes B, F, G and H, with  gluon $t$- or $u$-channel 
exchange, are infrared-divergent and thus, 
following Ref.~\cite{CKKR01}, are regularized with the inclusion 
of an effective gluon thermal mass in the gluon propagator. 
This method gives always positive definite values for the single
cross sections. The full results and the method how they are
obtained is described in the Appendix. 
In Table~\ref{table:channels},  we give for simplicity only the
first two leading terms in the expansion for $s/m_{\rm eff}^2 >> 1$.
However the logarithm in the approximate formulae of Table~\ref{table:channels}
gives unphysical negative value for $s< m_{\rm eff}^2$. Therefore in the numerical calculation
we have used the full formulae which are positive definite for all
values of $s$ listed in the Appendix.

The total cross-sections $\sigma_n$, where $n=A,\ldots,J$, can
be written as
\dis{
\sigma_n(s)=\frac{\alpha_s^3}{4\pi^2
    \fa^2}\bar{\sigma}_n (s),
\label{sigman}
}
where $\bar{\sigma}_n (s)$ denotes the cross-section averaged over
spins in the initial state and are given in Table~\ref{table:channels}.
The relevant multiplication factors are
also listed:
$n_s$ (the number of initial spin states), $n_F$ (the number of chiral
multiplets) and $\eta_i$ (the number density factor, which is $1$ for
bosons and $3/4$ for fermions).  We assume particles in thermal
equilibrium to have a (nearly) Maxwell-Boltzmann distribution,
proportional to $\eta_i $, and neglect Fermi blocking or Bose-Einstein
enhancement factors, which are close to one at these temperatures. We
restrict ourselves to temperatures above the freeze-out temperature of
SM superpartners involved, so that the approximation of thermal
equilibrium is always satisfied.\footnote{At lower temperatures,
  superpartner number densities, apart from the NLSP one, drop down to
  zero and the NTP regime is reached.}  Finally, in
Table~\ref{table:channels} the group theory factors $f^{abc}$ and
$T^a_{jk}$ of the gauge group $SU(N)$ satisfy the relations
$\sum_{a,b,c}|f^{abc}|^2=N(N^2-1)$ and
$\sum_a\sum_{jk}|T^a_{jk}|^2=(N^2-1)/2$.

Next we move to include contributions from the $SU(2)_L$ and $U(1)_Y$
gauge interactions.
The relevant axino-gaugino-gauge boson and axino-gaugino-sfermion-sfermion
interaction terms, in view of~\eq{eq:Leff}, are given by
\dis{
{\mathcal L}^{\rm eff}&=i\frac{\alpha_s}{16\pi \fa}\overline{\axino}\gamma_5[\gamma^\mu,\gamma^\nu]\tilde{g}^b
G^b_{\mu\nu} +\frac{\alpha_s}{4\pi \fa}\overline{\axino}\tilde{g}^a
\sum_{\tilde{q}}g_s \tilde{q}^*T^a\tilde{q}\\
&+i\frac{\alpha_2C_{aWW}}{16\pi \fa}\overline{\axino}\gamma_5[\gamma^\mu,\gamma^\nu]\tilde{W}^b
W^b_{\mu\nu}\\
&+\frac{\alpha_2}{4\pi \fa}\overline{\axino}\tilde{W}^a
\sum_{\tilde{f}_D}g_2 \tilde{f}_D^*T^a\tilde{f}_D\\
 &+ i\frac{\alpha_YC_{aYY}}{16\pi \fa}\overline{\axino} \gamma_5[\gamma^\mu,\gamma^\nu] \tilde{Y}Y_{\mu\nu}\\
&+\frac{\alpha_Y}{4\pi \fa}\overline{\axino}\tilde{Y}
\sum_{\tilde{f}}g_Y \tilde{f}^*Q_Y\tilde{f},
\label{eq:Laxino}
}
where the terms proportional to $\alpha_2$ come from the $SU(2)_L$ and the
ones proportional to $\alpha_Y$ from the $U(1)_Y$ gauge groups.
$C_{aWW}$ and $C_{aYY}$ are the model-dependent couplings for
the $SU(2)_L$ and $U(1)_Y$ gauge group
axino-gaugino-gauge boson anomaly interactions, which is defined after the standard
normalization of $\fa$, as in the first line for $SU(3)$, as stated below
\eq{Leff}.  Here $\alpha_2$, $\tilde{W}$, $W_{\mu\nu}$ and $\alpha_Y$,
$\tilde{Y}$, $Y_{\mu\nu}$ are the gauge coupling, the gaugino field
and the field strength of $SU(2)_L$ and $U(1)_Y$ gauge groups,
respectively. $\tilde{f_D}$ represents the sfermions of the
$SU(2)$-doublet and $\tilde{f}$ are the sfermions carrying the
$U(1)_Y$ charge.

\begin{table}
\begin{center}
\begin{tabular}{|c||c|l|c|c|c|} \hline
n & Process & \makebox[30mm][c]{$\sigmabar_n$} \hfill & \footnotesize{$n_s$}
& $n_F$ & \footnotesize{$\eta_1\eta_2$}\\
\hline &&&&&\\ [-1.3em]\hline
A & $ g^a + g^b \ra \tilde a + \tilde g^c $ &
$\frac{1}{24}|f^{abc}|^2$ & 4 & 1 & 1
\\ [0.2em]\hline
B & $ g^a + \tilde g^b \ra \tilde a + g^c $ &
\footnotesize{$\frac{1}{4}|f^{abc}|^2\left[\logsoms-\frac{7}{4}\right]$} & 4 & 1 &
$\frac{3}{4}$
\\ [0.2em]\hline
C & $ g^a + \tilde q_k  \ra \tilde a + q_j $ &
$\frac{1}{8}|T^{a}_{jk}|^2$ & 2 & \footnotesize{$2N_F$} & 1
\\ [0.2em]\hline
D & $ g^a + q_k  \ra \tilde a + \tilde q_j $ &
$\frac{1}{32}|T^{a}_{jk}|^2$ & 4 & $N_F$ &$\frac{3}{4}$
\\ [0.2em]\hline
E & $ \tilde q_j + q_k  \ra \tilde a + g^a $ &
$\frac{1}{16}|T^{a}_{jk}|^2$ & 2 & $N_F$ &$\frac{3}{4}$
\\ [0.2em]\hline
F & $ \tilde g^a + \tilde g^b \ra \tilde a + \tilde g^c $ &
\footnotesize{$\frac{1}{2}|f^{abc}|^2\left[\logsoms-\frac{23}{12}\right] $} & 4 & 1 &
$\frac{3}{4}\frac{3}{4}$
\\ [0.2em]\hline
G & $ \tilde g^a + q_k \ra \tilde a + q_j $ &
$\frac{1}{4}|T^{a}_{jk}|^2\left[\logsoms-2\right]$ & 4 & $N_F$ &
$\frac{3}{4}\frac{3}{4}$
\\ [0.2em]\hline
H & $ \tilde g^a + \tilde q_k \ra \tilde a + \tilde q_j $ &
\footnotesize{$\frac{1}{4}|T^{a}_{jk}|^2\left[\logsoms-\frac{7}{4}\right]\ast$} & 2 &
\footnotesize{$2N_F$} &$\frac{3}{4}$
\\ [0.2em]\hline
I & $ q_k + {\bar q_j} \ra \tilde a + \tilde g^a $ &
$\frac{1}{24}|T^{a}_{jk}|^2$ & 4 & \footnotesize{$\frac12N_F$} &
$\frac{3}{4}\frac{3}{4}$
\\ [0.2em]\hline
J & $ \tilde q_k + \tilde q_j^\ast \ra \tilde a + \tilde g^a $ & $\frac{1}{6}|T^{a}_{jk}|^2~\ast$ & 1 &
$N_F$ & 1
\\ [0.2em]\hline
\end{tabular}\caption{
The cross-section for each axino thermal-production
channel involving strong interactions. The particle masses are
neglected except for the plasmon mass $m_{\rm eff}$. The H and J
  entries with an asterisk in the third column are changed due to including the
  missing term and cross-sections or $n_F$ in the others processes
  (A,B,D,E,F, and I) are corrected from those of Ref.~\cite{CKKR01}.
  The logarithm in the approximate formulae  in this Table
gives unphysical negative value for $s< m_{\rm eff}^2$. Therefore in the numerical calculation
we used the full formulae which are positive definite for all values of $s$.
  The full cross sections for the processes 
B, F, G and H are given in the Appendix.}
\label{table:channels}
\end{center}
\end{table}

We start from Table~\ref{table:channels} and replace
quark triplets with $SU(2)_L$-doublets with corresponding group
factors. For the abelian $U(1)_Y$ factor, the processes A, B, and F
vanish and we can replace $|T^a_{jk}|^2$ with the square of the
corresponding hypercharges. Finally, we use $N_F=(12,14,11)$
to count the matter multiplets charged under the
MSSM gauge groups $(SU(3)$, $SU(2)_L$ and $U(1)_Y$, respectively.
We include above the SUSY breaking scale the full 1-loop 
MSSM running of the gauge couplings and gaugino masses.

The second term for each gauge group in~\eq{eq:Laxino} also generates
three-body gaugino decays into an axino and two sfermions, assuming
that the gauginos are heavy enough.  The three-body decay rate of the
gluino is given by
\dis{
\Gamma(\gluino^a \rightarrow \axino \tilde q_j
\tilde{q}^*_k)=\left(\frac{\alpha_s^2\mgluino^3}{128\pi^3\fa^2}
\right)\left(\frac{\alpha_s}{16\pi}{|T^a_{jk}|^2}\right) G\left(\frac{\msquark^2}{\mgluino^2}\right),
}
where  $\mgluino$ denotes the gluino mass,
\dis{
G(x)=&\frac23\sqrt{1-4x}(1+5x-6x^2)\\
&-4x(1-2x+2x^2)\log\left[\frac{1+\sqrt{1-4x}}{2x}-2 \right],
}
and the mass of the axino has been neglected.
However, the three-body decay is suppressed by an additional power of
the gauge coupling constant and is kinematically allowed only when the 
gluino mass is larger than the sum of the two final-state squark masses.
Therefore  gluino three-body decay through the second term in~\eq{eq:Laxino} is
subdominant to the two-body decay.

As stated in Ref.~\cite{CRS02}, an effective dimension-4 axino-quark-squark
coupling can be generated at a loop level also in the KSVZ model.
Here we take into account this effective Yukawa interaction, which
appears at a
two-loop level in the KSVZ models and a tree level (with a tiny mixing)
in the DFSZ models~\cite{CRS02},
\dis{
{\mathcal L}_{\axino\psi\tilde{\psi}}=g_{\rm
  eff}^{L/R}\tilde{\psi}^{J/R}_j \bar{\psi}_jP_{R/L}\gamma^5\tilde{a},
\label{Ldim4}
}
where $\psi_j$ and $\tilde{\psi}_j$ denote the SM fermions and their
superpartners.

In the KSVZ class of models, the effective coupling comes predominantly from the
logarithmically divergent part of the gluon-gluino-quark loop term and is
proportional to  $\mgluino$~\cite{CRS02},
\dis{
g_{\rm eff}^{L/R}\simeq \mp\frac{\alpha_s^2}{\sqrt2
  \pi^2}\frac{\mgluino}{\fa}\log\left(\frac{\fa}{\mgluino} \right).
\label{ggq-loop}
}
Subleading terms have not yet been computed and may give a correction
of the order of $20-30$\%, in analogy with what has recently been
obtained  in Ref.~\cite{Wyler09} for the effective tau-stau-axino coupling.

In the DFSZ models there exists also a tree-level axino-quark-squark
coupling which is proportional to the mass of the quark~\cite{Chun:2011zd},
coming from the $c_2$ interaction term in~\eq{eq:DFSZc2all},
\begin{equation}
g_{\rm eff}^{L/R}= \mp i \frac{m_q}{\fa}\frac{v_{d,u}^2}{v^2}=
\mp i \frac{m_q}{\fa} \left\{
\begin{array}{c}
\cos^2\beta \cr
\sin^2\beta,
\end{array} \right.
\label{ggq-tree}
\end{equation}
where the upper row relates to the up-type quarks and the lower row to
the down-type quarks. These tree-level couplings are always smaller
than the one-loop ones for the light generations, but not for the
third one.  In fact for the top quark, the tree-level coupling
dominates if $\tanb \lsim 4$ for the gluino mass of $700\gev$, while
the bottom quark tree-level coupling dominates if $\tanb \gsim 1$
for the same choice of the gluino mass.
Note that, at low reheating temperatures, only the top-stop-axino
coupling is important for axino thermal production. This is because
the lighter stop is usually the lightest colored superpartner and
remains in equilibrium to rather low temperatures, even below the EWSB scale. Similarly, there
exists an effective tau-stau-axino vertex, which was first obtained
in Ref.~\cite{CRRS04} and more recently re-derived in Ref.~\cite{Wyler09} in a
full two-loop computation. This coupling is smaller and not important
for thermal production, but it is important for the non-thermal
production when the stau is the NLSP.

In the DFSZ models, there is also a tree-level axino-Higgs-higgsino
coupling~\cite{Bae:2011jb}; compare the second term in \eq{eq:DFZSsuperpotential}, \dis{
  g_{\rm eff}=\frac{4\mu}{\sqrt2 \fa}, } where $\mu\sim
f_sV_a^2/\mplanck$ and $\fa=2V_a$. It contributes to axino TP through
higgsino decays in thermal equilibrium, and can be comparable to that
from squark decays through \eq{ggq-loop} and \eq{ggq-tree}, or even
become much larger if $\mu$ is larger than the top quark mass.
The axino production due to this coupling in DFSZ models
has been recently investigated also in  Ref.~\cite{Chun:2011zd}.
Note that the Yukawa coupling and the axino-higgsino mixing
contained in the neutralino mass matrix also give rise to
additional scattering channels contributing to the axino production,
but we will neglect them here since such dimension-4 scatterings are usually less important
than the decays~\cite{CRS02}.
\begin{figure}[!t]
  \begin{center}
  \begin{tabular}{c}
   \includegraphics[width=0.6\textwidth]{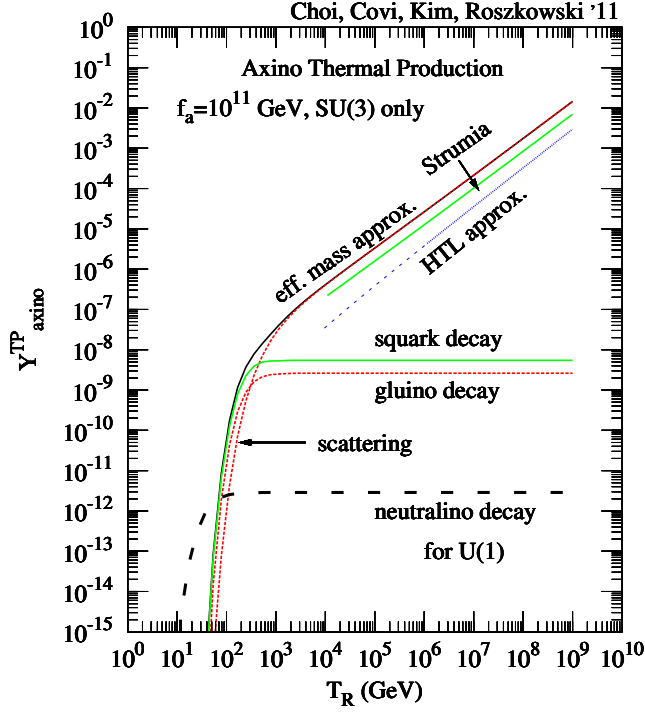}
   \end{tabular}
  \end{center}
  \caption{Thermal axino yield $Y_\axino^{\rm TP}$ as a function of
    the reheating temperature $\treh$ from strong interactions using
    the effective mass approximation (black). We use the
    representative values of $\fa=10^{11}\gev$ and
    $m_{\squark}=\mgluino=1\tev$. For comparison, we also show the HTL
    approximation  (dotted blue/dark grey) and that of Strumia (green/light grey).
    We also denote the yield from squark (solid green/light grey) and 
    gluino decay (dotted red), as well as out-of-equilibrium bino-like 
    neutralino decay (dashed black). Here we used the interactions in 
    \eq{eq:Laxino} and \eq{ggq-loop} for the KSVZ  model.
    We use the same definition of reheating temperature
      in the instantaneous reheating approximation for the three methods.}
\label{fig:YTR}
\end{figure}
\begin{figure}[t]
  \begin{center}
  \begin{tabular}{c}
   \includegraphics[width=0.6\textwidth]{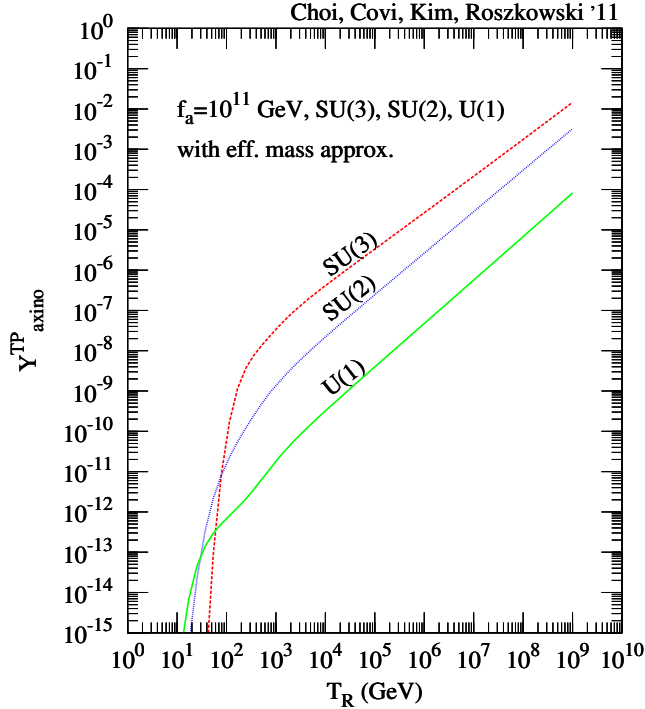}
      \end{tabular}
  \end{center}
  \caption{Thermal axino yield $Y_\axino^{\rm TP}$ as a function of
$\treh$ from each of the SM gauge groups. Here, we have used $C_{aWW}=C_{aYY}=1$.
 The lines at high $T_R$ are not perfectly parallel due to the running
 of the gauge couplings, which affects the $SU(3) $ yield more
 strongly and in the opposite direction than the other gauge groups.}
\label{fig:YTR_all}
\end{figure}
\begin{figure}[tbh!]
  \begin{center}
  \begin{tabular}{c}
   \includegraphics[width=0.6\textwidth]{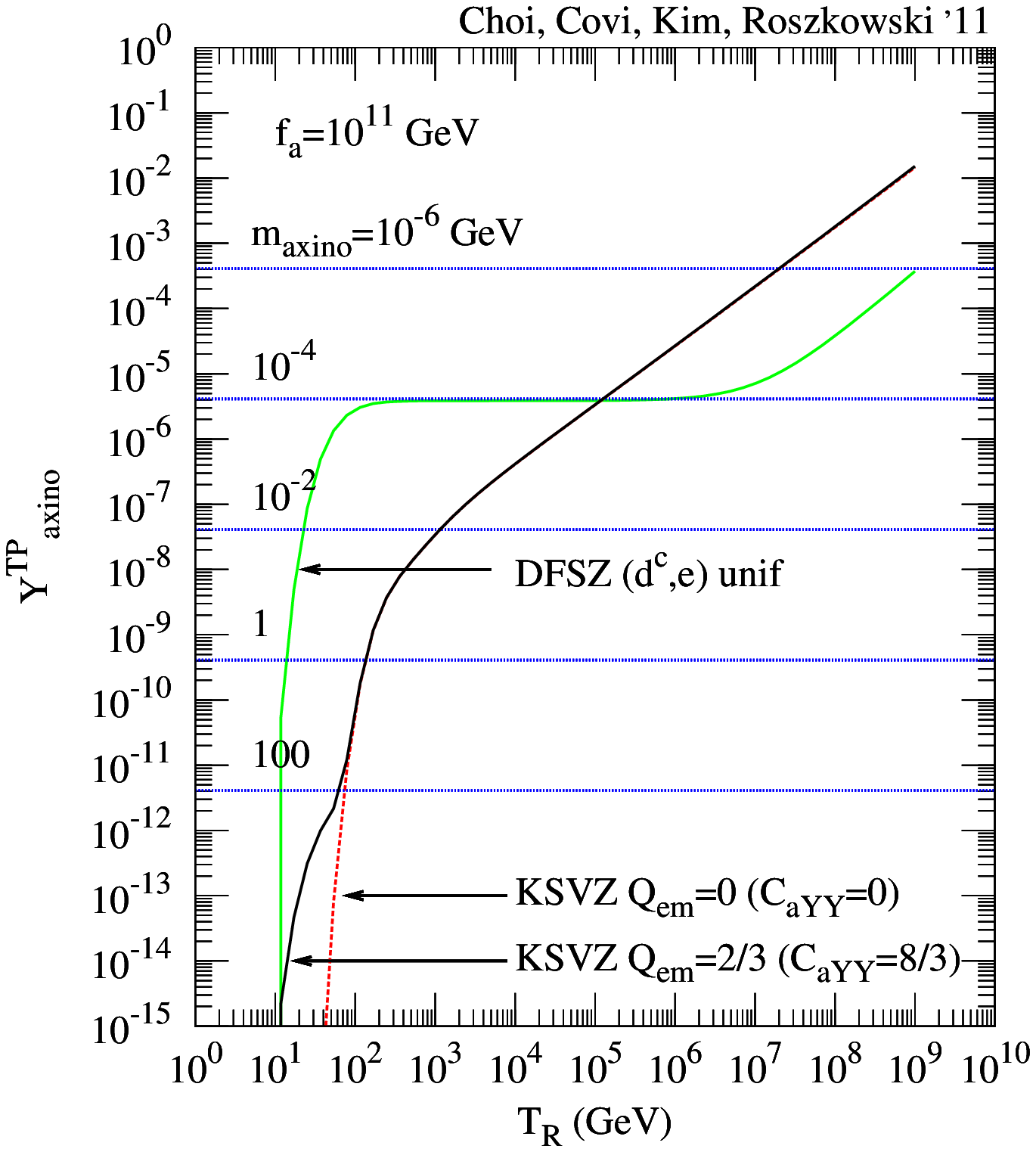}
   \end{tabular}
  \end{center}
  \caption{Thermal axino yield $Y_\axino^{\rm TP}$ as a function of
    $\treh$ for two specific KSVZ models: $Q_{\rm em}=0$ ($C_{aYY}=0$) and
    $Q_{em}=2/3$ ($C_{aYY}=8/3$), and for a DFSZ model with the $(d^c,e)$
    unification~\cite{Kim:1998va}, for which we used $\mu=200\gev$ and the
    higgsino mass $m_{\tilde{h}}=200\gev$. The horizontal lines show the
    values of axino mass for which the corresponding axino abundance  gives
    the correct DM relic density.}
\label{fig:YTR_KSVZ}
\end{figure}

We have evaluated the thermal production of axinos numerically and present
the results in Figures~\ref{fig:YTR} and ~\ref{fig:YTR_all} for representative
values of $\fa=10^{11}\gev$ and $\msquark=\mgluino=1\tev$. 
We do not consider here the dependence on masses, however see Ref.~\cite{CRS02}.
 For different values
of $\fa$ the curves move up or down proportional to $\fa^2$.
In Fig.~\ref{fig:YTR}, we show the axino yield $Y$ (where $Y\equiv
n/s$ is the ratio of the number density to the entropy density) from
strong interaction in the KSVZ model.
Our result obtained with the effective mass approximation is
shown with the solid black line.
Compared to the previous plot in Ref.~\cite{CKKR01}, the inclusion of the
squark decay changes the plot at low reheating temperature, while
the other new squark interactions do not have any noticeable
effect. There is a factor 3 difference in the abundance at high reheating temperature compared to that 
 in Figure 2 of Ref.~\cite{CKKR01}, which was a numerical error at
 that time and was corrected later.  
For comparison, the axino yield from scatterings using the
HTL approximation~\cite{Steffen04} is plotted with the blue (dashed) line
and Strumia's result~\cite{Strumia10} is shown with the green line.

For $\treh\gtrsim 10^4\gev$  the axino abundance using the effective
mass approximation increases consistently with that of Strumia. We found
that the difference between the two prescriptions is of order a factor of three.
 In principle we could reabsorb this difference in the definition of 
the effective gluon mass at high temperature, which in our scheme 
cannot be determined self-consistently. With this tuning we could 
match the perturbative result at high temperature.
On the other hand, doing this would require a gluon thermal mass smaller 
than the expression $ \sim  g T$ used in our calculations. Hence we prefer 
instead to consider this factor as an estimate of the theoretical 
error of using the effective mass approximation at high
reheating temperatures. We assume that this error 
does not increase for reheating temperatures less than $10^4\gev$
and we apply the effective mass approximation to the DFSZ model.

For lower temperatures the contributions from the decays of squarks
and gluinos in thermal plasma, which were not included
in Ref.~\cite{Strumia10}, start playing some role. We mark those in
Fig.~\ref{fig:YTR} with a green solid and red dashed curves,
respectively.  It is known that, at reheating temperatures above
superpartner masses scattering diagrams involving dimension-4
operators are usually subdominant relative to those coming from
dimension-5 operators and to decay terms, and are negligible.
Also the decays do not give significant contribution to the TP of axinos,
apart from very low $\treh$~\cite{CRS02}, and this is confirmed in
Fig.~\ref{fig:YTR}.
While all the above contributions are generated
by strong interactions only, for comparison, we show also (as a 
black dashed line) the relative contribution from an
out-of-equilibrium bino-like neutralino decay to an
axino and a photon originally considered in Ref.~\cite{CKR00}.
It is clear that NTP is only important at very low $\treh$, well below
squark or gluino masses.

In Fig.~\ref{fig:YTR_all}, we show a contribution to the axino yield
in thermal production from each SM gauge group interaction. Here we
set the coefficients $C_{aWW}=1$ and $C_{aYY}=1$ as a normalization.
As shown in the figure, the contributions from scatterings due to
$SU(2)_L$ and $U(1)_Y$ couplings (blue dotted and green solid lines,
respectively) are significantly suppressed compared to that of
$SU(3)_c$ (red dashed), by a factor of 10 or more. This is because the
interaction between axinos and gauge bosons are proportional to a
gauge coupling-squared so that the cross-section is $\sigma\propto
\alpha^3$.  Thus it would be only for very large (and perhaps
unnatural) values of the effective couplings $C_{aWW}, C_{aYY} $ that
these channels could become comparable to the QCD contribution. 
To give an order of magnitude estimate of these effects,
we included the SU(2) and U(1) contributions with a normalized
value in figure 2. For different values of $ C_{aWW} $ and
$C_{aYY} $ the curves move up and down.
We note that in general SUSY breaking effects in the leptonic
sector may bring a modification of the couplings here considered.
The situation here is different from the case of gravitino production
since the interactions of the gravitino to the three gauge groups are
of the same order: the spin-3/2 gravitino component couples in fact
universally, while the goldstino component proportionally to the
gaugino masses. 
We therefore conclude that the QCD contribution is
strongly dominant in the KVSZ models and so the axino production at
high $\treh$ is practically model independent as long as the number of
heavy PQ charged states can be absorbed into the definition of $\fa$.

However, at low $\treh$ the thermal production from scatterings
becomes strongly suppressed by the Boltzmann factor. In the region
where $\treh \lsim 100-1000\gev$, axino production due to the decays
of gaugino, squarks or neutralinos become important. Actually,
the lightest neutralino decay via $U(1)_Y$ couplings becomes dominant in
the very low reheating temperature regime since the number density of
the heavier colored particles becomes strongly suppressed by the
Boltzmann factor there.  On the other hand the neutralinos, depending
on their composition and the supersymmetric spectrum, can freeze-out
with a still substantial number and then give rise also to non-thermal
axino production, as we have seen in Fig.~\ref{fig:YTR}. This
contribution to the axino yield is usually more important than the one
due to neutralino decays in equilibrium, which is proportional to
$C_{aYY} $ and typically below $10^{-12}$.

For the case of the DFSZ instead, the role of the QCD interaction
is played by the $SU(2)$ interaction and the dominant
decay term above the EW symmetry breaking is the higgsino one
instead of the squark one.

Our results for the total thermal production yield for both KVSZ
and DFSZ type of models can be seen in 
Figure~\ref{fig:YTR_KSVZ}.\footnote{A similar figure for the DFSZ model
 is given in Ref.~\cite{Bae:2011jb}.}
There we show the KSVZ model with different values of the
$ C_{aYY}$ coupling. In the case of non-zero $C_{aYY}$ the
contribution from neutralino decay in equilibrium can be
clearly seen for $ T_{R} \sim 10 $ GeV and it is very suppressed.
Moreover we give also the yield for the DFSZ model in solid green, for
$\mu=200\gev$ and the higgsino mass $m_{\tilde{h}}=200\gev$.
We can see that even for relatively small $\mu$, axino production
from higgsino decay  dominates over the one from the anomaly terms
for reheating temperature  $\treh\lesssim 10^6\gev$. The importance of axino-Higgsino-Higgss
coupling in the DFSZ model was
recently discussed  in \cite{Chun:2011zd} and our result is consistent
with that analysis.  The abundance is so large
that the CDM density can be reached with an axino  mass as small as
$100\kev$, independently of the reheating temperature.\footnote{
Such general effect due to decaying particles in equilibrium has been
recently called ``freeze-in'' in Ref.~\cite{Hall:2009bx} and discussed for
the axino in Ref.~\cite{Cheung:2011mg}. The freeze-in mechanism was
included already in gluino or squark decays to axinos in the
plasma in Ref.~\cite{CRS02}.}
In this range of reheating temperature, the axino production
  from decay dominates that from scatterings. Therefore the use of 
the effective mass approximation or another IR screening prescription 
in the scattering process is irrelevant to the axino production in 
the DFSZ model in the range of reheating temperature where the decay dominates.
For higher $\treh$, the $SU(2)_L$ anomaly term starts dominating
and the abundance is proportional to $ T_R$ as in the KVSZ case,
but with a smaller coefficient.
In the same figure, we also mark horizontal lines
corresponding to the axino mass giving the correct DM relic density
for the given relic abundance of $Y_{\axino}$.

\begin{figure}[t]
  \begin{center}
  \begin{tabular}{c}
   \includegraphics[width=0.6\textwidth]{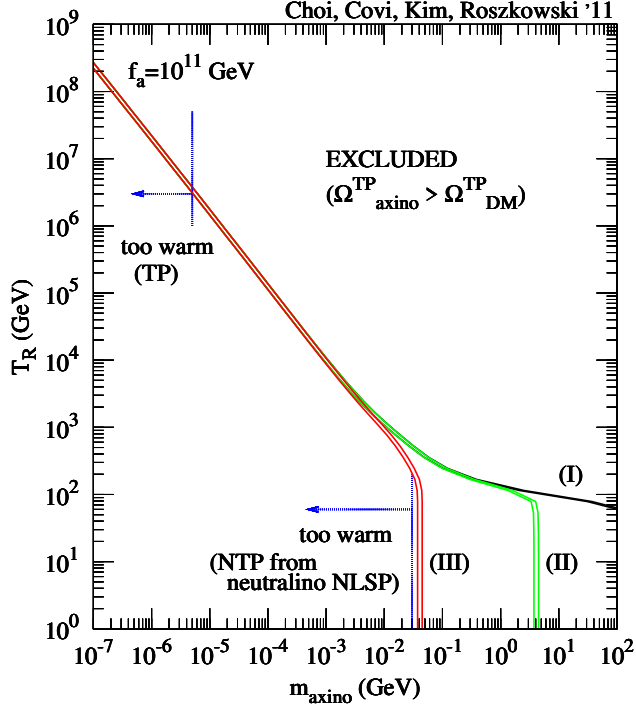}
   \end{tabular}
  \end{center}
  \caption{$\treh$ versus $\maxino$ for $\fa=10^{11}\gev$ in the KSVZ
    models. The bands
    inside like curves correspond to a correct relic density of DM
    axino with both TP and NTP included. To parametrize the
    non-thermal production of axinos we used $Y_{\rm NLSP}=0$ (I),
    $10^{-10}$ (II), and $10^{-8}$ (III). The upper right-hand area of
    the plot is excluded because of the overabundance of axinos. The
    regions disallowed by structure formation are marked with vertical
    blue dashed lines and arrows for, respectively, TP
    ($\maxino\lsim5\kev$, see text below \eq{v0}) and NTP ($\maxino\lsim30\mev$, with a
    neutralino NLSP).  }
\label{fig:TR_ma}
\end{figure}

\begin{figure}[t]
  \begin{center}
  \begin{tabular}{c}
   \includegraphics[width=0.6\textwidth]{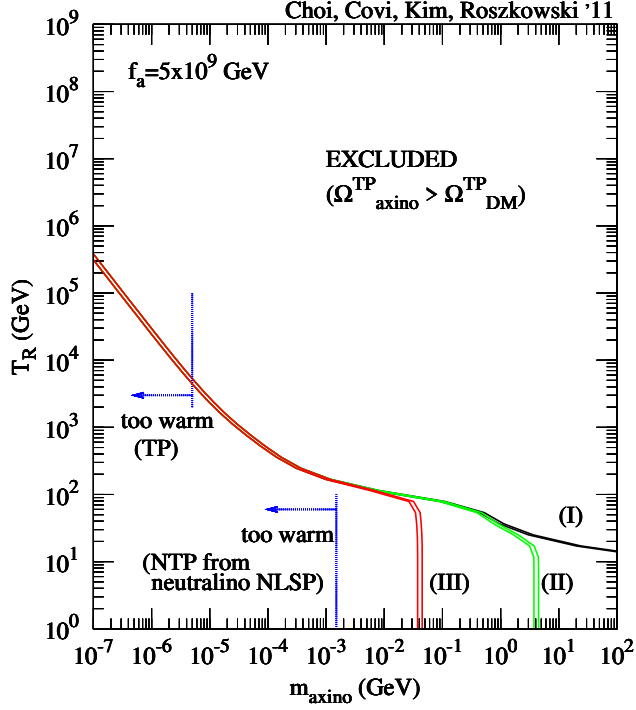}
   \end{tabular}
  \end{center}
 \caption{The same as Fig.~\ref{fig:TR_ma} but with the PQ scale $\fa=5\times10^9\gev$.  }
\label{fig:TR_ma2}
\end{figure}

\subsection{Non-thermal production}

As stated above, axinos can be produced non-thermally in NLSP decays after they have
frozen out of equilibrium. This NTP mechanism is dominant for reheating temperatures
below the mass of the gluino and squarks~\cite{CKR00,CKKR01}.
In this case, the axino abundance is independent of the reheating
temperature as long as the temperature is high enough for the NLSP
to thermalize before freeze-out. Axino relic density from
NTP is simply given by
\dis{
\omegaantp=\frac{\maxino}{\mnlsp}\Omega_{\rm NLSP}
\simeq2.7\times10^{10}\bfrac{\maxino}{100\gev}Y_{\rm NLSP}.
}
Clearly, in order to produce a substantial NTP population of axinos,
the NLSP must itself have an energy density larger than the present
DM density.

\begin{figure}[t]
  \begin{center}
  \begin{tabular}{c}
   \includegraphics[width=0.6\textwidth]{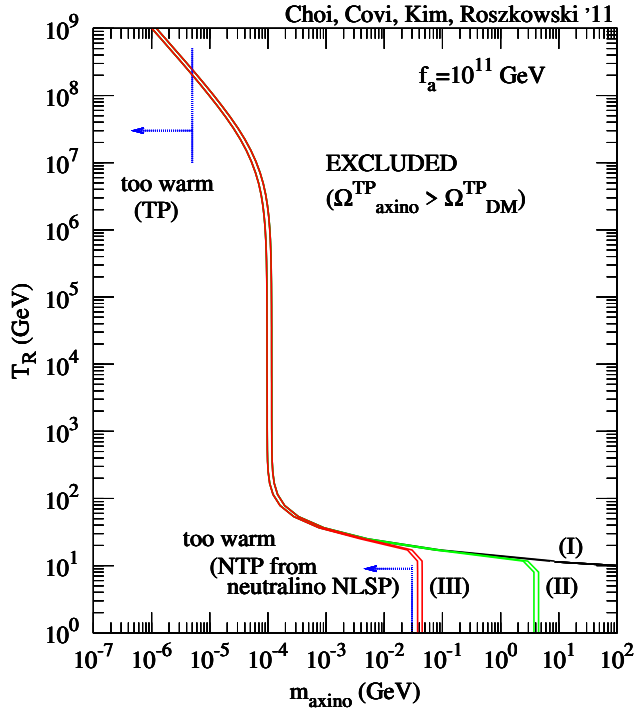}
   \end{tabular}
  \end{center}
  \caption{The same as Fig.~\ref{fig:TR_ma} but with DFSZ model used in Fig.~\ref{fig:YTR_KSVZ}.   }
\label{fig:TR_ma_DFSZ}
\end{figure}

\begin{figure}[t]
  \begin{center}
  \begin{tabular}{c}
   \includegraphics[width=0.6\textwidth]{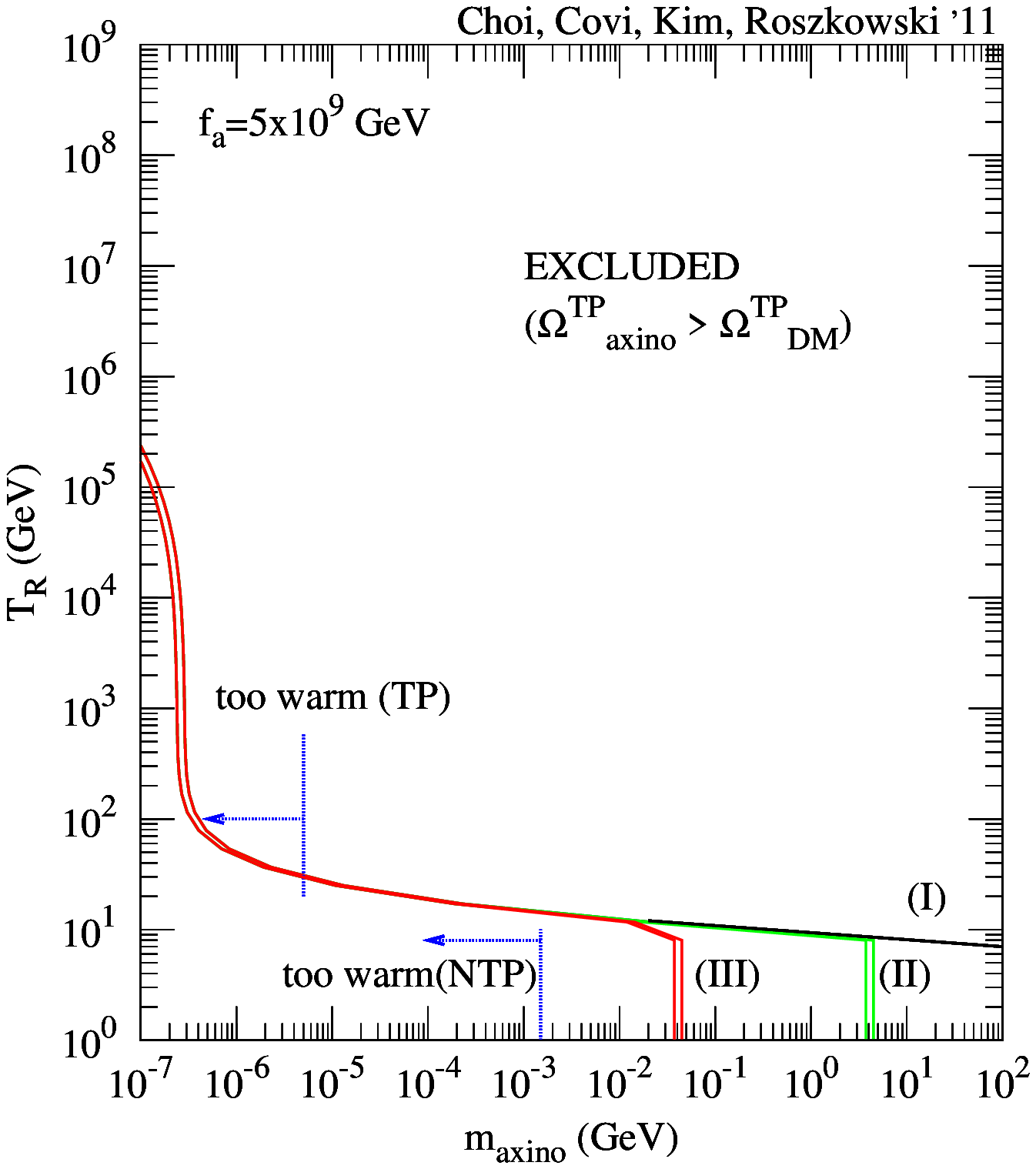}
   \end{tabular}
  \end{center}
 \caption{The same as Fig.~\ref{fig:TR_ma_DFSZ} but with the PQ scale $\fa=5\times10^9\gev$.  }
\label{fig:TR_ma_DFSZ2}
\end{figure}

To see if such production is sufficient to give a dominant DM component, we need to know the
yield of NLSPs after they have frozen out of the thermal plasma.
For the neutralino NLSP yield, relevant processes include pair-annihilation
and co-annihilation with the charginos, next-to-lightest neutralinos and
sleptons. For the  stau NLSP, the yield is determined by the
stau-stau annihilation and stau-neutralino co-annihilation processes.
A typical relic abundance is
\dis{
Y_\chi \simeq (1\sim10)\times10^{-12}\bfrac{m_\chi}{100\gev},
\label{eq:Yneutralino}
}
for a bino-dominated neutralino, and
\dis{
Y_\stau \simeq 0.01\times10^{-12}\bfrac{m_\stau}{100\gev},
\label{eq:Ystau}
}
for the stau. Note that in the latter case the NTP can produce
sufficient axino abundance to explain the whole DM density
only for stau masses above $ 1.9\tev $, which may thermalize only
at accordingly higher temperatures.

These two choices for the NLSP were considered in the Constrained
Minimal Supersymmetric Standard Model (CMSSM) in Ref.~\cite{CRRS04},
for $\fa < 10^{12}\gev $, for which even the stau lifetime is of order
$1 \mbox{s}$, or less. Recently, the case of stau NLSP, including
four-body hadronic decays, was discussed in Ref.~\cite{Wyler09} also
for larger values of $ \fa$.

In conclusion, for neutralino NLSP, which decays mainly into an axino
and a photon or a $Z$-boson, the NTP production is usually more
efficient.  For the stau NLSP, which can decay to an axino and a
tau-lepton through a coupling of the type given in \eq{Ldim4}, the
contribution is smaller, but can still be substantial.

Regarding other NLSPs, colored relics (or even a wino-like neutralino
if it is lighter than $~1.8\tev$) usually remain in thermal
equilibrium so long that their number density after freeze-out becomes
negligible and therefore cannot produce any substantial axino
population after freeze-out~\cite{Berger:2008ti, Covi:2009bk}.

\section{Cosmological constraints}\label{sec:CosBounds}

\subsection{The relic density of dark matter}

For the total axino DM relic density, we apply the $3\sigma$ range
derived from WMAP-7 data~\cite{Komatsu:2010fb}
\dis{ 0.109 < \abunda
  <0.113.
\label{eq:wmap7_3sigma} }
This produces a stripe in the parameter space and also
plays a role of the upper bound on the relic density when there are
additional DM components, e.g. the axion.

This can be seen in Figs.~\ref{fig:TR_ma} and~\ref{fig:TR_ma2}, where
we present the reheating temperature versus the axino mass plane for
$\fa=10^{11}\gev$ and $5\times10^9\gev$, respectively. We have
included both the thermal and the non-thermal production contributions of
axinos, the latter assuming $Y_{\rm NLSP}=0$ (black solid) $10^{-10}$
(green solid) and $10^{-8}$ (red dash), denoted also as (I), (II), and
(III), respectively. A typical stau and neutralino yield after freezout
will lie between (I) and (III). A correct relic
density of axinos, in the range given by \eq{eq:wmap7_3sigma},
corresponds to the thin bands between like curves. The parameter space
above the curves is excluded as giving too much relic abundance.
Similar figures in the DFSZ model are presented in Figs.~\ref{fig:TR_ma_DFSZ}
and~\ref{fig:TR_ma_DFSZ2}, which can be compared to the figures in Ref.~\cite{Bae:2011jb}.

\subsection{Nucleosynthesis}

An injection of high energy electromagnetic and hadronic particles
during or after Big Bang Nucleosynthesis (BBN) epoch may disrupt the
abundances of light elements. For axino DM, the lifetime of the NLSP,
such as the neutralino or the stau, is typically around $1\second$, or less,
and therefore constraints from the BBN are weak. However, for longer
lifetimes such constraints become
important~\cite{Jedamzik:2004er,kkm04}.  This leads to an upper bound
on the decay products of the NLSP for a given NLSP lifetime.  For the
stau NLSP, a bound state with ${}^4{\rm He}$ severely constrains its
lifetime to be less than roughly $5\times
10^3\sec$~\cite{Pospelov:2006sc} (although in specific cases with
gravitino as DM, this can be up to an order of magnitude
larger~\cite{bcjr09}). However for the parameters considered in
our study, i.e. $ \fa < 10^{12}\gev $, the BBN constraint can be
avoided due to the small lifetime of the NLSP. For larger values of
$\fa$, on the other hand, non-trivial constraints arise, especially
for the stau NLSP, as recently discussed in Ref.~\cite{Wyler09}.

\subsection{Structure formation}

The density perturbation due to axino population is suppressed at scales below
their free-streaming length. When the axino mass is larger than
${\mathcal O}(\kev)$, thermally produced axinos become
cold~\cite{CKKR01}. However, the non-thermal population of axinos from
NLSP decays can still have a large velocity dispersion and can be too
warm. Lyman-$\alpha$~\cite{Boyarsky:2008xj} and
reionization data~\cite{Jedamzik:2005sx} give a bound on the velocity of
the WDM component and its fraction in the DM density. A recent
analysis using the SDSS Lyman-$\alpha$ data~\cite{Boyarsky:2008xj}
leads to an upper limit on the average velocity, $\langle v \rangle_{\rm
  WDM}< 0.013\, \rm{km/s}$ for pure WDM, or otherwise in the case of
mixed cold/warm DM the WDM fraction is constrained to be $\Omega_{\rm
  WDM}/\Omega_{\rm DM} < 0.35$ in the larger velocity region.

The present velocity of thermally produced axinos is given
by~\cite{Barkana:2001gr}
\dis{
v_0\simeq 0.065\, {\rm km/s}\, \bfrac{1\kev}{\maxino}.
\label{v0}
}
Therefore the above Lyman-$\alpha$ data implies $\maxino \gtrsim 5\kev$
for TP axinos. This lower bound is marked in Figs.~\ref{fig:TR_ma}
and~\ref{fig:TR_ma2} with a vertical blue dashed line and an arrow.

The free-streaming velocity of axinos produced
non-thermally can be obtained from the lifetime of the NLSP and the
mass relations~\cite{Jedamzik:2005sx,CKKR01,CRRS04}.  For the
bino-like neutralino NLSP with $C_{\rm aYY}=8/3$, we find
\dis{
v_0\simeq 0.4 {\rm km/s}
\bfrac{\maxino}{1\mev}^{-1}\bfrac{m_\chi}{100\gev}^{-1/2}\bfrac{\fa}{10^{11}\gev},
}
and for the stau NLSP,
\dis{
v_0\simeq 2 {\rm km/s}\,&
\bfrac{\maxino}{1\mev}^{-1}\bfrac{m_\stau}{100\gev}^{1/2}\\
&\times\bfrac{m_\chi}{100\gev}^{-1}\bfrac{\fa}{10^{11}\gev}.
}

For NTP axinos and for the neutralino NLSP we find a lower limit on the axino
mass $\maxino \gtrsim 30 \mev$ and $1.5\mev$ when
  $\fa=10^{11}\gev$ and $5\times10^9\gev$, respectively. These limits
  are marked in Figs.~\ref{fig:TR_ma} and ~\ref{fig:TR_ma2} with a
vertical blue dashed line and an arrow. For the stau NLSP the analogous lower
bounds are $\maxino \gtrsim 150\mev$ and $7.5\mev $, respectively.
We stress that these bounds apply solely if the population of axinos
produced through NTP is substantial ($\gsim20-30\%$).

%
\begin{figure}[t]
  \begin{center}
  \begin{tabular}{c}
   \includegraphics[width=0.6\textwidth]{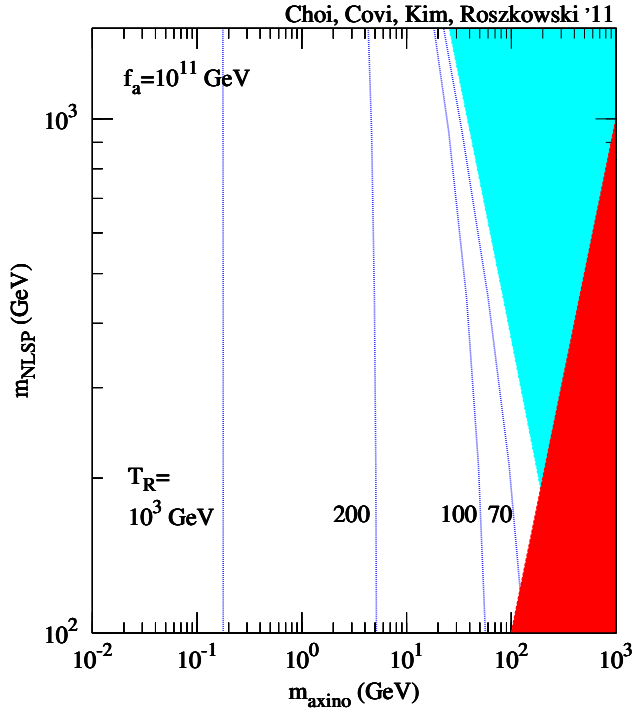}
   \end{tabular}
  \end{center}
 \caption{Contours of constant reheating temperature in the
    NLSP--axino mass plane. Here we have assumed $Y_{\rm
      NLSP}=10^{-12}\left({\mnlsp}/{100\gev}\right)$, typical of
    neutralino NLSP, and taken $\fa=10^{11}\gev$. The cyan wedge in the
    upper right-hand corner is excluded by the overdensity of DM,
    while in the red wedge below it the axino is not the LSP.    }
\label{fig:ms_ma_A}
\end{figure}
\begin{figure}[t]
  \begin{center}
  \begin{tabular}{c}
   \includegraphics[width=0.6\textwidth]{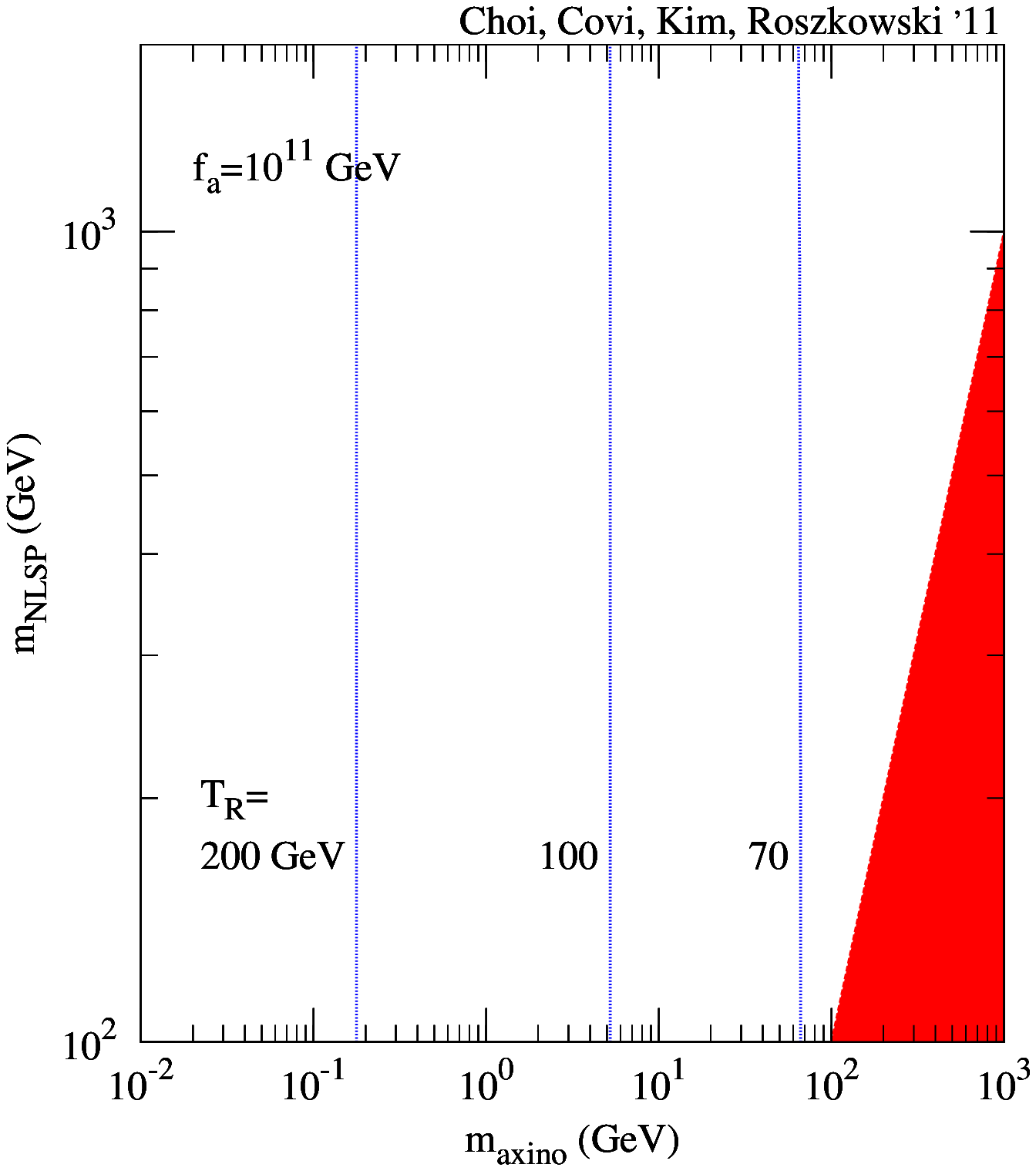}
   \end{tabular}
  \end{center}
 \caption{The same as Fig.~\ref{fig:ms_ma_A} except for
$Y_{\rm NLSP}=10^{-14}\left({\mnlsp}/{100\gev}\right)$, typical of  stau NLSP.
 }
\label{fig:ms_ma_B}
\end{figure}

In Figs.~\ref{fig:ms_ma_A} and ~\ref{fig:ms_ma_B}, we show contours of
the reheating temperature in the plane spanned by the NLSP and the
axino mass.~\footnote{Similar figures are shown in Ref.~\cite{Wyler09}.} In Fig.~\ref{fig:ms_ma_A} we have fixed $\fa=10^{11}\gev$
and assumed $Y_{\rm NLSP}=10^{-12}\left({\mnlsp}/{100\gev}\right)$,
which is a typical value for  bino-like neutralino NLSP. The cyan wedge in the
upper right-hand corner is excluded by the overdensity of DM, while in
the red wedge below it the axino is not the LSP. In
Fig.~\ref{fig:ms_ma_B} instead $Y_{\rm
  NLSP}=10^{-14}\left({\mnlsp}/{100\gev}\right)$ has been assumed,
typical of stau NLSP.
So long as the TP dominates, the curves of constant $\treh $ remain
vertical and practically independent of $\maxino $, while as soon as
the NTP becomes important, $\maxino $ dependence arises leading to
non-vertical curves. For example, with a bino-like neutralino NLSP of
$100\gev $, as in Fig.~\ref{fig:ms_ma_A}, the NTP contributes only up
to 10 \% of the axino LSP DM density.  For stau NLSP with small
abundance $Y\lsim 10^{-13}$, as given in~\eq{eq:Ystau}, NTP is
always subdominant so that the contours remain vertical. In this case
the bounds coming from the free-streaming velocity of the NTP axinos
are absent.

\section{Conclusion}\label{sec:Concl}

We have performed an updated analysis of the relic axino production,
taking into account some new calculations that have appeared after the
initial study~\cite{CKR00,CKKR01}, especially~\cite{Steffen04,Strumia10}
for the thermal part, and compared them with our own results
to explore the question of uncertainty and model-dependence.

We have found that the uncertainty has not really decreased after the latest
calculation~\cite{Strumia10}. This is probably not surprising since
the QCD coupling is large and the convergence of the perturbative
series is quite slow. Comparing the different results, we estimate the 
uncertainty in the relic density of axinos produced thermally to be 
still of order a factor of 10 or so at $T_{R}\sim 10^4 $ GeV, and to 
that one has to add also some possible (unknown) contributions due to 
non-perturbative effects.
Our result lies above both estimates given in
Refs.~\cite{Steffen04,Strumia10}, and this seems natural since we 
included subleading terms in $m_{\rm eff}^2/s$, which do indeed
increase the cross-sections for  single channels ensuring their 
positivity in the whole range of integration, even in the limit
of very large gauge coupling.

Regarding the model dependence, our conclusions are more optimistic:
in the KSVZ-type the QCD anomaly term strongly dominates the axino
thermal production mechanism, apart from the case of small reheating
temperatures where sparticle decay contributions start playing the dominant
role. The inclusion of the additional anomaly couplings is completely
negligible, apart from unnaturally large values of the coefficients
$C_{aWW}$ and $C_{aYY}$.
Instead for DFSZ-type models, the Yukawa interaction dominates
around the weak scale and can give the main production mechanism of
axinos making it independent of the reheating temperature.
Therefore the axino abundance is free from the uncertainty in the method
used for the IR-divergence.
At large temperatures, it is the EW anomaly term that dominates, giving
 a lower abundance than in the KSVZ case.

 For both models, also the non-thermal production via NLSP decay can
 produce the required axino DM density, if the NLSP decouples with a
 sufficiently large abundance. But for this mechanism to dominate, the
 reheating temperature has to be very low and the axino and NLSP
 masses not too hierarchical.  We find that, interestingly enough, a
 light bino NLSP decaying out of equilibrium can still produce the
 whole DM density at the cost of a very low reheating temperature (but
 sufficiently high for NLSP thermalization). For the stau case
 instead, it is quite unlikely that the NTP can dominate, unless the
 stau NLSP yield after freeze-out is unusually large.

\medskip

During the final stages of completion of this work, the analysis 
Ref.~\cite{Bae:2011jb} appeared which discusses in details  axino couplings
and finds a non-trivial momentum dependence in the one-particle irreducible
one-loop axino-gluon-gluino couplings. The coefficient $ C_{1PI} $ of
these interactions is suppressed when the external particle momentum
is much larger than the mass $M$ of the PQ-charged fermions in the loop.
Due to this effect, the authors claim a suppression of order $ M^2/T^2 $
of the axino production from the dimension-5 operators for the DFSZ
case and for extremely small KSVZ quark masses (with Yukawa 
couplings less than $10^{-5}$, \ie for the heavy quark mass less than 
$10^6 \gev$ for $f_a\simeq 10^{11}\gev$). 

We investigated such suppression by inserting their $ C_{1PI} $
coupling in the relevant diagrams and we obtain different suppressions
depending on the type of Feynman graph and a strong dependence on
the IR regulator contained there, in most cases the gluon thermal mass.
In particular, we find no suppression at all for the $t$-channel gluon 
exchange for vanishing gluon mass.
Since graphs with the one loop $ C_{1PI} $ coupling and a gluon thermal mass
insertion arise at lowest order at two loops, probably a full
investigation of the two-loop diagrams in thermal field theory is
needed to resolve this issue.

\begin{figure}[!t]
  \begin{center}
  \begin{tabular}{c}
   \includegraphics[width=0.6\textwidth]{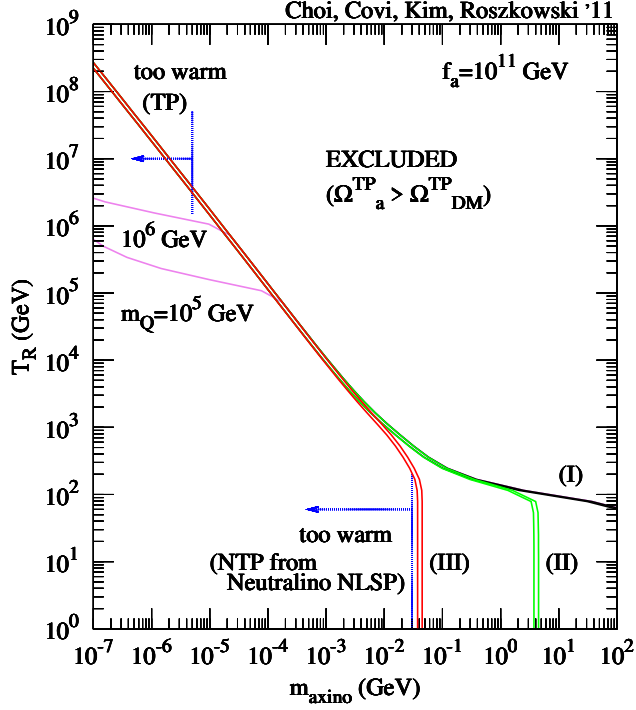}
   \end{tabular}
  \end{center}
  \caption{The same as Fig.~\ref{fig:TR_ma} but showing as well
           in magenta the yield suppression found in
           Ref.~\cite{Bae:2011jb} for different values of the heavy
           quark masses.   }
\label{TR-mQ}
\end{figure}

Note in any case that even without suppression, the DFSZ axino 
production is dominated by the decay term up to temperatures of the 
order of $ 10^{6-7} $ GeV. For the KVSZ case, we show in Fig.~\ref{TR-mQ}
as violet lines how the yield changes according to
Ref.~\cite{Bae:2011jb} for small heavy quark masses, $m_Q=10^6 \gev$
and $m_Q=10^5 \gev$. The {\em Cold} DM axino, on which our present study is 
based, is practically not affected.
For large fermion masses in the loop, $ M > T $, our anomalous
couplings coincide fully with the $ C_{1PI} $ in
Ref.~\cite{Bae:2011jb} and our results are in perfect agreement.

\medskip
\acknowledgments{The authors would like to thank the Galileo Galilei
  Institute and its Program "Dark Matter: Its nature, origins and
  prospects for detection" for hospitality when this work was
  initiated. KYC is in part supported by the Korea Research Foundation
  Grant funded from the Korean Government (KRF-2008-341-C00008 and
  No. 2011-0011083)).  KYC acknowledges the Max Planck Society (MPG),
  the Korea Ministry of Education, Science and Technology (MEST),
  Gyeongsangbuk-Do and Pohang City for the support of the Independent
  Junior Research Group at the Asia Pacific Center for Theoretical
  Physics (APCTP). LC would like to thank the CERN Theory Division for
  hospitality in the context of the TH-Institute DMUH'11 (19-29 July 2011)
  during the completion of this work.  JEK is supported in
  part by the National Research Foundation (NRF) grant funded by the
  Korean Government (MEST) (No. 2005-0093841). 
  LR is partially supported by the Welcome Programme of the Foundation for Polish
 Science, by the Lancaster, Manchester and Sheffield Consortium for Fundamental Physics under STFC grant ST/J0000418/1, and by the EC 6th Framework Programme MRTN-CT-2006-035505.}


\appendix

\section{Appendix}

In this Appendix, we present the computation of the axino relic
  density in detail.
The time evolution of the axino number density, $n_\axino$, is
described by the Boltzmann equation:
\dis{
\frac{d n_\axino}{dt} + 3H n_\axino = \sum_{i,j} \VEV{\sigma(i+j\rightarrow \axino + \cdots)v_{\rm rel}}n_i n_j + \sum_i \VEV{\Gamma(i\rightarrow \axino +\cdots)}n_i. \label{Friedmann}
}
Here $H$ is the Hubble parameter, $H(T)=\sqrt{(\pi^2 g_*)/(90\mplanck^2)} T^2$, where $g_*$ is the number of effective massless degrees of freedom. The first term in the r.h.s. of \eq{Friedmann}
is the axino production from the scattering process of particles $i$ and $j$  in the thermal bath with cross section
$\sigma(i+j\rightarrow \axino + \cdots)$ and relative velocity $v_{\rm rel}$. $n_i$ is the $i$-th particle's number density in the thermal bath. The second term in the r.h.s. of \eq{Friedmann} is the axino production from the decay of $i$ particle which are in thermal bath with the decay rate $\Gamma(i\rightarrow \axino +\cdots)$. $\VEV{\cdots}$ denotes the thermal averaging including averaging over initial spins and summing over the final spins.
Here we neglect the inverse processes  since the axinos are decoupled from the thermal bath
and their number density is very small.

Using the axino yield defined as
\dis{
Y \equiv \frac{n_\axino}{s},
}
where $s=(2\pi^2/45)g_{s*} T^3$ is the entropy density, and normally $g_{s*}=g_*$ in the early Universe,
the solution of the Boltzmann equation is  
\dis{
Y_\axino = \sum_{i,j}Y^{\rm scat}_{i,j} + \sum_iY_i^{\rm dec},
}
where
\dis{
&Y^{\rm scat}_{i,j}=\int^{\treh}_0 dT \frac{ \VEV{\sigma(i+j\rightarrow \axino + \cdots)v_{\rm rel}}n_i n_j  }{sHT},\\ 
&Y_i^{\rm dec}=\int^{\treh}_0 dT \frac{\VEV{\Gamma(i\rightarrow \axino +\cdots)}n_i}{sHT}.
}
The explicit formulae for the thermal average are given in Ref.~\cite{Choi:1999xm}.

The relevant 2-body scattering cross sections are summarized in 
table~\ref{table:channels}, keeping in mind that the physical
cross-sections are
 \eq{sigman}
\dis{
\sigma_n(s)=\frac{\alpha_s^3}{4\pi^2
    \fa^2}\bar{\sigma}_n (s).
}
Then $\bar{\sigma}_n (s)$ is given from the matrix element, $M_n$, by
\dis{
\bar{\sigma}_n (s)=\frac{\pi\fa^2}{4\alpha_s^3}\frac{1}{ s^2}\int^0_{-s}|M_n|^2dt.\label{sigmabarn}
}
Among the processes,  B,F,G and H have an infrared (IR) divergence due 
to the massless gluon exchange in the $t$- or $u$-channel. To
regularize this IR divergence, we introduce, only in the $t$ or
$u$-channel gluon propagators, an effective thermal gluon mass 
$m_{\rm eff}^2=g^2T^2$,
which is generated by plasma effects~\cite{Ellis:1984eq,CKKR01}.
Then we obtain finite and always positive cross sections. 
For example, the squared matrix element for process B with using massless gluon propagator is 
\dis{
|M_B|^2=\frac{-4(2s^2+ 2ts +t^2)}{t}|f^{abc}|^2.
} However introducing effective thermal gluon mass in the $t$- or $u$-channel, 
it changes to
\dis{
|M_B|^2=\frac{-2t(3\mg^4-6\mg^2t+4s^2+4ts+2t^2)}{(t-\mg^2)^2}|f^{abc}|^2,
}
which is positive definite by definition. Then $\bar{\sigma}_B (s)$ is obtained after integration over $t$ as given in the \eq{sigmabarn}.
Here we list the full cross-sections $\bar{\sigma}_n$ for these processes.  
\begin{enumerate}
\item Process B:
\dis{
\frac{|f^{abc}|^2}{16 s^2 \left(\mg^2+s\right)}\left[s \left(3 \mg^4-5 s \mg^2-7 s^2\right)+\left(-3 \mg^6+5 s \mg^4+12 s^2 \mg^2+4 s^3\right) \log \left((\mg^2+s)/\mg^2\right)\right].\label{F_full}
}
\item Process F: We include the factor $1/2$ for the identical initial particles,
\dis{
&\frac{(1/2)|f^{abc}|^2}{12s^2 \left(2 \mg^2+s\right)}
\left[-s \left(24 \mg^4+58 s \mg^2+23
s^2\right)\right.\\
&\qquad \qquad\left.+12 \left(2 \mg^6+6s \mg^4+4s^2\mg^2+s^3\right)
\log \left((\mg^2+s)/\mg^2\right)\right].
}
\item Process G:
\dis{
\frac{|T^{a}_{jk}|^2}{4s}\left[-2 s+\left(2 \mg^2+s\right) \log \left((\mg^2+s)/\mg^2\right)\right].
}
\item Process H:
\dis{
\frac{|T^{a}_{jk}|^2}{16 s^2 \left(\mg^2+s\right)}\left[-s \left(3 \mg^4+9 s \mg^2+7 s^2\right)+\left(3 \mg^6+11 s \mg^4+12 s^2 \mg^2+4 s^3\right) \log \left((\mg^2+s)/\mg^2\right)\right].\label{H_full}
}
\end{enumerate}

These formulas give back the expressions in Table~\ref{table:channels}, 
in the limit of large $s/\mg$. We mention here that those expressions in Table~\ref{table:channels}
provide unphysical result for small  $s/\mg$ and give negative contribution to the thermal averages.
In fact substituting $s\simeq T^2, \mg^2 \sim g^2 T^2 $, one finds 
e.g. for the B process
\begin{equation}
\bar \sigma_B \propto \log[1/g^2(T)] - \frac{7}{4} < 0
\quad\quad \mbox{for} \quad\quad g^2(T) > 0.17
\end{equation}
and this holds for all the interesting region up to the GUT scale
where $g^2 \sim 0.5 $. So even if the total thermal cross-section 
may result positive, it is lower than the real one due to
the negative contribution of these IR-divergent channels.
On the other hand the full expressions above from \eq{F_full} to \eq{H_full} are positive for any value
of $s$. We can see that easily in the limit of large $\mg^2/s$ because those reduce to:
\begin{enumerate}
\item Process B:
\dis{
\frac{3|f^{abc}|^2}{32} + {\cal O} \left( \frac{s}{\mg^2} \right)
}
\item Process F: 
\dis{
\frac{|f^{abc}|^2}{24} +  {\cal O} \left( \frac{s^3}{\mg^6} \right)
}
\item Process G:
\dis{
\frac{|T^{a}_{jk}|^2}{24} \frac{s^2}{\mg^4} +  {\cal O} \left( \frac{s^3}{\mg^6} \right)
}
\item Process H:
\dis{
\frac{|T^{a}_{jk}|^2}{32} +  {\cal O} \left( \frac{s}{\mg^2} \right)
}
\end{enumerate}
so they have the correct asymptotic behaviour corresponding to
screening and physical decoupling of the intermediate gluon channels 
for large gluon mass.~\footnote{Note that the B and H channels also 
contain diagrams without intermediate gluon and therefore remain
finite also in the limit of infinite gluon mass.}
So with these formulas the single cross-sections are always
positive and the thermal average larger than in the previous
estimates.

In our numerical computations we use the full formulae above for the
IR divergent processes, while for the other processes,  
$A, C, D, E, I, J$, we use the expressions are given in Table~\ref{table:channels}.


\def\prp#1#2#3{Phys.\ Rep.\ {\bf #1} (#3) #2}
\def\rmp#1#2#3{Rev. Mod. Phys.\ {\bf #1} (#3) #2}
\def\anrnp#1#2#3{Annu. Rev. Nucl.
Part. Sci.\ {\bf #1} (#3) #2}
\def\npb#1#2#3{Nucl.\ Phys.\ {\bf B#1} (#3) #2}
\def\plb#1#2#3{Phys.\ Lett.\ {\bf B#1} (#3) #2}
\def\prd#1#2#3{Phys.\ Rev.\ {\bf D#1}, #2 (#3)}
\def\prl#1#2#3{Phys.\ Rev.\ Lett.\ {\bf #1} (#3) #2}
\def\jhep#1#2#3{J. High Energy Phys.\ {\bf #1} (#3) #2}
\def\jcap#1#2#3{J. Cosm. and Astropart. Phys.\ {\bf #1} (#3) #2}
\def\zp#1#2#3{Z.\ Phys.\ {\bf #1} (#3) #2}
\def\epjc#1#2#3{Euro. Phys. J.\ {\bf #1} (#3) #2}
\def\ijmp#1#2#3{Int.\ J.\ Mod.\ Phys.\ {\bf #1} (#3) #2}
\def\mpl#1#2#3{Mod.\ Phys.\ Lett.\ {\bf #1} (#3) #2}
\def\apj#1#2#3{Astrophys.\ J.\ {\bf #1} (#3) #2}
\def\nat#1#2#3{Nature\ {\bf #1} (#3) #2}
\def\sjnp#1#2#3{Sov.\ J.\ Nucl.\ Phys.\ {\bf #1} (#3) #2}
\def\apj#1#2#3{Astrophys.\ J.\ {\bf #1} (#3) #2}
\def\ijmp#1#2#3{Int.\ J.\ Mod.\ Phys.\ {\bf #1} (#3) #2}
\def\apph#1#2#3{Astropart.\ Phys.\ {\bf B#1}, #2 (#3)}
\def\mnras#1#2#3{Mon.\ Not.\ R.\ Astron.\ Soc.\ {\bf #1} (#3) #2}
\def\apjs#1#2#3{Astrophys.\ J.\ Supp.\ {\bf #1} (#3) #2}
\def\aipcp#1#2#3{AIP Conf.\ Proc.\ {\bf #1} (#3) #2}


\end{document}